\documentstyle[orosztwo]{article}

\begin{document}

\title{Optical Observations of 
GRO J1655-40 in Quiescence I:
A Precise Mass for the Black Hole Primary}

\author{Jerome A. Orosz\altaffilmark{1,2} and Charles D. 
Bailyn\altaffilmark{1,3}}
\affil{Department of Astronomy, Yale University, P. O. Box 208101,
New Haven, CT 06520-8101  \\ 
E-Mail:  orosz@astro.yale.edu, bailyn@astro.yale.edu}

\altaffiltext{1}{Visiting Astronomer at Cerro Tololo Inter-American 
Observatory 
(CTIO), which is operated by the Association of Universities for Research 
in Astronomy Inc., under contract with the National Science
Foundation.} 
\altaffiltext{2}{Present Address:  Department of Astronomy \& Astrophysics,
The Pennsylvania State University, 525 Davey Laboratory, University Park, PA 
16802}
\altaffiltext{3}{National Young Investigator}

\begin{abstract}

We report photometric and spectroscopic observations of the
black hole binary GRO J1655-40 in complete
quiescence.  In contrast to the 1995 photometry, the light curves from 1996
are almost completely dominated by ellipsoidal modulations from
the secondary star.  Model fits to the light curves, 
which take into account the temperature
profile of the accretion disk and eclipse effects, yield an inclination
of $i=69.50\pm 0.08\deg$ and a mass ratio of $Q= 
M_1/M_2=2.99\pm 0.08$.  
The precision of our determinations of $i$ and $Q$ allow us to
determine the black hole mass to an accuracy of $\approx 4\%$
($M_1=7.02\pm 0.22\,M_{\sun}$).  The secondary star's mass is
$M_2=2.34\pm 0.12\,M_{\sun}$.
The position of the secondary
on the Hertzsprung-Russell diagram is consistent
with that of a $\approx 2.3\,M_{\sun}$ star which has evolved off the
main sequence and is halfway to the start of the giant branch.  
Using the new spectra we present an
improved value of the spectroscopic period ($P=2\fd 62157\pm 0\fd 00015$),
radial  velocity semiamplitude 
($K=228.2\pm 2.2$~km~s$^{-1}$), and
mass function ($f(M)=3.24\pm 0.09\,M_{\sun}$).
Based on the new spectra of the source and
spectra of several MK spectral type standards,
we classify the secondary star as F3IV to F6IV.
Evolutionary models suggest an
average
mass transfer rate for such a system of 
$\dot M_2=3.4\times
10^{-9}\,M_{\sun}~{\rm yr}^{-1}$ $=2.16\times 10^{17}$~g~s$^{-1}$,
which is much larger
than the average mass transfer rates implied in the other six transient
black hole systems, but still barely below the critical mass transfer
rate required for stability.
\end{abstract}


\section{Introduction}

A new bright
X-ray source (designated GRO J1655-40) was discovered July 27, 1994
with the Burst and Transient Source Experiment (BATSE) on board the
{\em Compton Gamma Ray Observatory} 
(Zhang et al.\ 1994\markcite{zh94}).
The optical counterpart and the radio counterpart were identified
soon after the announcement of the discovery by the BATSE team
(Bailyn et al.\ 1995a\markcite{ba95a}; 
Campbell-Wilson \& Hunstead 1994\markcite{cwh94}).
About three weeks after the initial hard X-ray outburst, radio
jets with apparent superluminal velocities were observed emerging from the
source 
(Tingay et al.\ 1995\markcite{ti95}; 
Hjellming \& Rupen 1995\markcite{hr95}), 
making
GRO J1655-40 only the second source in our Galaxy observed to have
superluminal jets and the first one to be optically identified.

Bailyn et al.\ (1995b)\markcite{bomr95b} 
established the spectroscopic period 
($2\fd 601\pm 0\fd 027$) and the mass function  
($3.16\pm 0.15\,M_{\sun}$) of the system using spectroscopic observations
made during late April and early May, 1995.
The photometric observations by 
Bailyn et al.\ (1995b)\markcite{bomr95b}
from early 1995 showed that GRO J1655-40 is an eclipsing binary.
The light curves from March and April, 1995 all have deep triangular-shaped
minima near the spectroscopic phase 0.75 (caused by an eclipse of
the disk by the star) and much shallower minima near the spectroscopic
phase 0.25 (caused by an eclipse of the star by the disk).   These
observations confirmed 
earlier hints that GRO J1655-40 might be an eclipsing binary:
Bailyn et al.\ (1995a)\markcite{ba95a} 
observed a single eclipse-like event in the photometry
on the night of August 17, 1994, and the model of the kinematics of the
radio jets proposed by 
Hjellming \& Rupen (1995)\markcite{hr95} 
gives an inclination of $i=85\deg$.
In spite of the observations of optical eclipses, there has not yet
been an unambiguous observation of an X-ray eclipse
(e.g.\ 
Harmon et al.\ 1995a\markcite{ha95a}).  
The lack of X-ray eclipses and the presence
of optical eclipses puts a tight constraint on the inclination of the orbit.

Unlike most ``X-ray novae,'' GRO J1655-40 continued to have major outburst
events in hard X-rays long after its initial high-energy outburst
(see 
Harmon et al.\ 1995a\markcite{ha95a}).  
There was an outburst event late March,
1995 
(Wilson et al.\ 1995)\markcite{whzpf95} 
and another one starting 
late July, 1995 
(Harmon et al.\ 1995b\markcite{ha95b}).  
Not surprisingly, the $V$ magnitude of the source during 1995
was typically around $V\approx 16.5$ 
(e.g.\ Bailyn et al.\ 1995b)\markcite{bomr95b},
somewhat higher than the quiescent value of $V\approx 17.3$
(Bailyn et al.\ 1995a\markcite{ba95a}).  
Furthermore, the shapes of the light curves
from 1995 are complicated.  There are night-to-night brightness variations,
and the phases of some of the optical minima are not aligned precisely
with the spectroscopic phase 
(Bailyn et al.\ 1995b)\markcite{bomr95b}.  
X-ray heating and hot spots on an asymmetric accretion disk probably 
play a large role in explaining the complex light curves.  

After the July/August, 1995 hard X-ray outburst, the source 
finally settled into true X-ray quiescence.  From the period of late
August, 1995 to late April, 1996, the source was not detected by
BATSE 
(Robinson et al.\ 1996\markcite{ro96}).  
The ASCA X-ray satellite made several pointed observations in late March,
1996 as part of a
large multi-wavelength observing program 
(Robinson et al.\ 1996\markcite{ro96}).  
The soft X-ray luminosity was found to be quite low, with
$L_x\approx 2\times 10^{32}$~ergs~s$^{-1}$ (assuming a distance
of $d=3.2$~kpc---see 
Robinson et al.\ 1996\markcite{ro96}), 
which is about a factor
of 900 times {\em smaller} than the optical luminosity of
the secondary star  ($L_{2}\approx 47\,L_{\sun}$, see Section
6).   The extended period of X-ray quiescence ended late April, 1996
when the all sky monitor on the 
{\em Rossi X-ray Timing Explorer} satellite detected a sharp increase
in brightness in the soft X-rays 
(Remillard et al.\ 1996a)\markcite{re96}.
The source also brightened significantly in the optical and UV
(Horne et al.\ 1996)\markcite{ho96}, 
and in the radio wavelengths 
(Hunstead \& Campbell-Wilson 1996\markcite{cwh96};
Hjellming \& Rupen 1996\markcite{hr96}).

\begin{deluxetable}{lcc}
\tablewidth{0pt}
\tablecaption{Journal of Spectroscopic Observations} 
\tablehead
{                                                                       &
1995 April 30-May 4\tablenotemark{a,b}   &
1996 February 24-25\tablenotemark{c}
}
\startdataT
Telescope    &   CTIO 4.0 m    &  CTIO 1.5 m  \nl
Detector     &   Loral $3072\times 1024$ & Loral
$1200\times 800$  \nl
Grating      & KPGL \#3 ($527 \ell/{\rm mm}$)
&   KPGL \#3 ($527 \ell/{\rm mm}$)  \nl
Resolution   & 3.3~\AA\, (FWHM) & 4.0~\AA\, (FWHM) \nl
Wavelength Coverage  & 3850-7149~\AA & 4673-6657~\AA \nl
Number of Spectra   & 73  & 12 \nl  
\enddata
\tablenotetext{a}{Previously published in 
Bailyn et al.\ 1995b\markcite{bomr95b}.}
\tablenotetext{b}{The heliocentric Julian date (HJD) range:
$2\,449\,837.580$ to $2\,449\,841.915$}
\tablenotetext{c}{HJD range: $2\,450\,137.792$ to $2\,450\,138.890$}
\label{sco1tab1}
\end{deluxetable}

We have obtained
additional photometry and spectroscopy of GRO J1655-40
February and March, 1996. 
Much of the March data was taken as part of the larger multi-wavelength
program
(Robinson et al.\ 1996\markcite{ro96}). 
These 
data show that the mean $V$ magnitude is consistent with its pre-outburst
value and that
in the optical the system is dominated by light from the
secondary star.  Since the accretion disk contributes a small fraction 
of the light  and since GRO J1655-40 eclipses, we have
a unique opportunity to model the light
curves and obtain a reliable measure of the orbital inclination, thereby
leading for the first time a reliable mass for a black hole.
In Section 2 
we describe our quiescent photometric observations. We present an improved
spectroscopic ephemeris
in Section 3. In Section 4 we discuss the spectral classification of the 
secondary star.  In Section 5 we describe
models of the light curves.  Section
6 is the discussion section detailing 
the evolutionary status of the secondary star, the inclination of the radio
jet,
and the stability of the accretion.
A short summary of the paper is presented in Section 7.  We have also included
an Appendix to this paper which gives a detailed description of the
code used to model the light curves.

\section{Observations and Reductions}\label{mIIsec2}
 
The spectroscopic observations from April and May, 1995 published in
Bailyn et al.\ (1995b)\markcite{bomr95b} 
are summarized in Table \ref{sco1tab1}. 
Additional spectra were taken in photometric conditions
February 24-25, 1996 with the RC spectrograph
on the CTIO 1.5 meter telescope.   
The KPGL \#3 grating and the Loral
$1200\times 800$ CCD combination gave a dispersion of 1.68~\AA\, per
pixel.  A He-Ar lamp was observed repeatedly to give the wavelength
calibrations.  In addition to the 12 spectra of GRO J1655-40,
we also obtained the spectra of 45
different stars (most with two or more observations),
including several radial velocity standards, flux standards,  and
MK spectral type standards 
(Morgan \& Kennan 1973)\markcite{mk73}.  
 
\begin{deluxetable}{lcccc}
\tablewidth{0pt}
\tablecaption{Journal of Photometric Observations} 
\tablehead{
{\begin{tabular}{c}   UT Date  \\                       \end{tabular}}  &
 Telescope                                                              &
 Filter                                                                 &
 Detector                                                               &
{\begin{tabular}{c}  Number \\  of exposures            \end{tabular}}  }
\startdataT
1996 February 5,7-20\tablenotemark{a} & CTIO 0.9 m & V & Tek 2k \#3 &  24  \nl
1996 February 5,7-20 & CTIO 0.9 m & I & Tek 2k \#3 &  23  \nl
&                         &            &        &                \nl
1996 March 21-31\tablenotemark{b}     & CTIO 0.9 m & B & Tek 2k \#3 &  24  \nl
1996 March 21-31     & CTIO 0.9 m & V & Tek 2k \#3 &  157  \nl
1996 March 21-31     & CTIO 0.9 m & R & Tek 2k \#3 &  24  \nl
1996 March 21-31     & CTIO 0.9 m & I & Tek 2k \#3 &  150  \nl
&                         &            &        &                \nl
1996 April 1\tablenotemark{c}         & CTIO 0.9 m & B & Tek 2k \#3 &  2  \nl
1996 April 1         & CTIO 0.9 m & V & Tek 2k \#3 &  12  \nl
1996 April 1         & CTIO 0.9 m & R & Tek 2k \#3 &  2  \nl
1996 April 1         & CTIO 0.9 m & I & Tek 2k \#3 &  9  \nl
\enddata
\tablenotetext{a}{HJD range: $2\,450\,118.836-2\,450\,133.892$}
\tablenotetext{b}{HJD range: $2\,450\,163.711-2\,450\,173.895$}
\tablenotetext{c}{HJD range: $2\,450\,174.727-2\,450\,174.895$}
\label{sco1tab2}
\end{deluxetable}

Photometry of the source was obtained February, March, and April, 1996 with
the CTIO 0.9 meter telescope and Tek $2048\times 2046$ \#3 CCD (see Table
\ref{sco1tab2}).  All of the nights in February, 1996 were photometric
and about half of the nights in the March-April, 1996 run were photometric.
Standard IRAF tasks were used to process the images
to remove the electronic bias and to perform flatfielding corrections.
The programs DAOPHOT IIe   and DAOMASTER 
(Stetson 1987\markcite{st87}; 
Stetson, Davis, \& Crabtree 1991\markcite{sdc91}; 
Stetson 1992a\markcite{st92a},b) 
were used to compute the photometric time series of GRO J1655-40 and several
field comparison stars.  The errors in the instrumental magnitudes were
estimated by computing the standard deviations in the light curves of
several stable field stars.  The sizes of these errors were found to
be in reasonable agreement with the errors reported by the DAOMASTER
program.  We
have adopted the errors given by DAOMASTER
in the analysis presented below.
The instrumental magnitudes were transformed
to the standard system using previously calibrated field stars
(Bailyn et al.\ 1995a\markcite{ba95a}).  
All of the February, 1996 frames were usable.
Out of the all the frames taken in March and April, one frame in $B$,
two frames in $R$, and one frame in $I$ were not used in any of the analysis
because of cosmic ray hits on or very near the image of GRO J1655-40.

\begin{deluxetable}{lr}
\tablewidth{0pt}
\tablecaption{Orbital Parameters for GRO J1655-40}
\tablehead{
 parameter   &
 result  }
\startdataT
Orbital period, spectroscopic (days) & $2.62157\pm 0.00015$\nl
K velocity (km~s$^{-1}$) & $228.2\pm 2.2$\nl
$\gamma$ velocity (km~s$^{-1}$) & $-142.4\pm 1.6$\nl
$T_0$ (spectroscopic) (HJD 2\,440\,000+)  & $9\,839.0763\pm 0.0055$\nl
Mass function ($M_{\sun}$)  &  $3.24\pm 0.09$\nl
\enddata
\label{sco1tab3}
\end{deluxetable}

\begin{figure*}
\epsscale{1.05}
\plotone{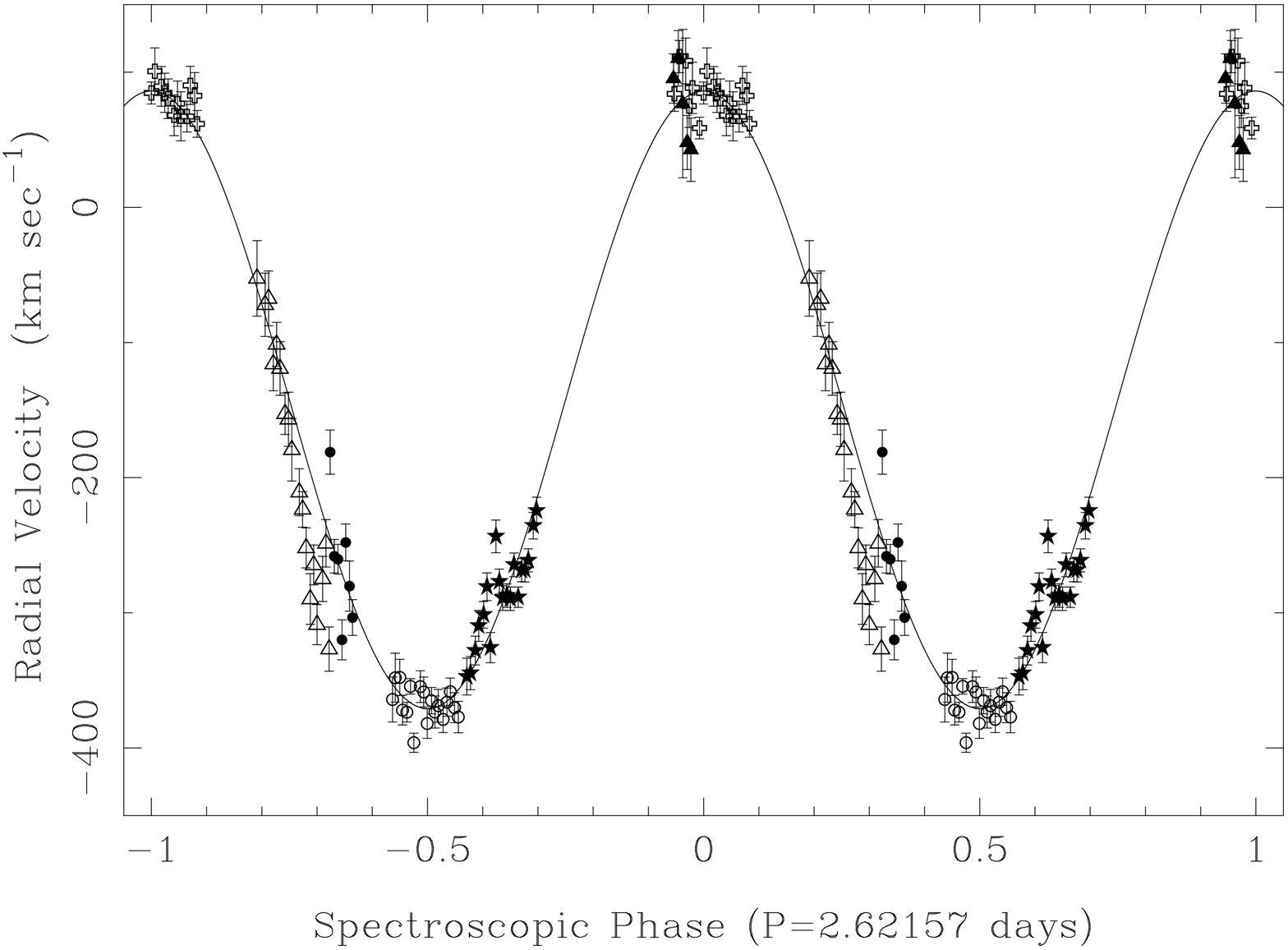} 
\caption{The folded radial velocities of GRO J1655-40 and the best fitting
sinusoid (see Table \protect\ref{sco1tab3}).
Each point has been plotted twice.
Open diamonds indicate data
from April 30, 1995; open triangles, filled
stars, and open crosses indicate data from
May 2, 3, and 4, 1995, respectively.  
The velocities from February 24, 1996 are indicated by the filled triangles,
and the filled circles indicate data from February 25, 1996.
Parts of this figure appeared in 
Bailyn et al.\ (1995b).}
\label{sco1fig1}
\end{figure*}

\section{A Refined Spectroscopic Ephemeris}

The 73 spectra from April 30-May 4, 1995 were rebinned to match the slightly
larger dispersion and smaller wavelength coverage of the 12 spectra
from 1996.  The radial velocities of the 85 spectra were found by
computing the cross-correlations 
(Tonry \& Davis 1979)\markcite{td79}
against a spectrum of 40 Leo
(=HD 89449), a radial velocity standard star of spectral type F6IV
(radial velocity $=+6.5\pm 0.5$~km~s$^{-1}$)
that was observed February, 1996.
The cross-correlations were computed over the wavelength region between
H$\beta$ and H$\alpha$, excluding the strong interstellar absorption
lines 
(see Bailyn et al.\ 1995a\markcite{ba95a}) 
and regions corrupted by bad pixels on
the CCDs.  Out of the 85 spectra, 84 yielded significant cross
correlations (`r' values greater than 4; see 
Tonry \& Davis 1979)\markcite{td79}.
The remaining spectrum had a huge cosmic ray hit and was not used
in any of the analysis presented below. A sinusoid fit, excluding
the May 2, 1995 data 
(when the star was partially eclipsed by the disk)
was performed giving the spectroscopic elements listed in
Table \ref{sco1tab3}.
The folded radial velocities and the best fitting sinusoid are shown in
Figure \ref{sco1fig1}.

We relaxed the assumption of a circular orbit and attempted to fit the
radial velocities to an eccentric orbit.  We applied the 
Lucy \& Sweeney test (1971)\markcite{ls71}
to check the significance of the derived eccentricity ($e=0.057\pm 0.020$).
The significance of the eccentric orbit fit is much lower than 5\%, indicating
the eccentric orbit fit is no improvement over the circular orbit fit to
the radial velocities. 

\begin{figure*}[p]
\epsscale{1.75}
\plotone{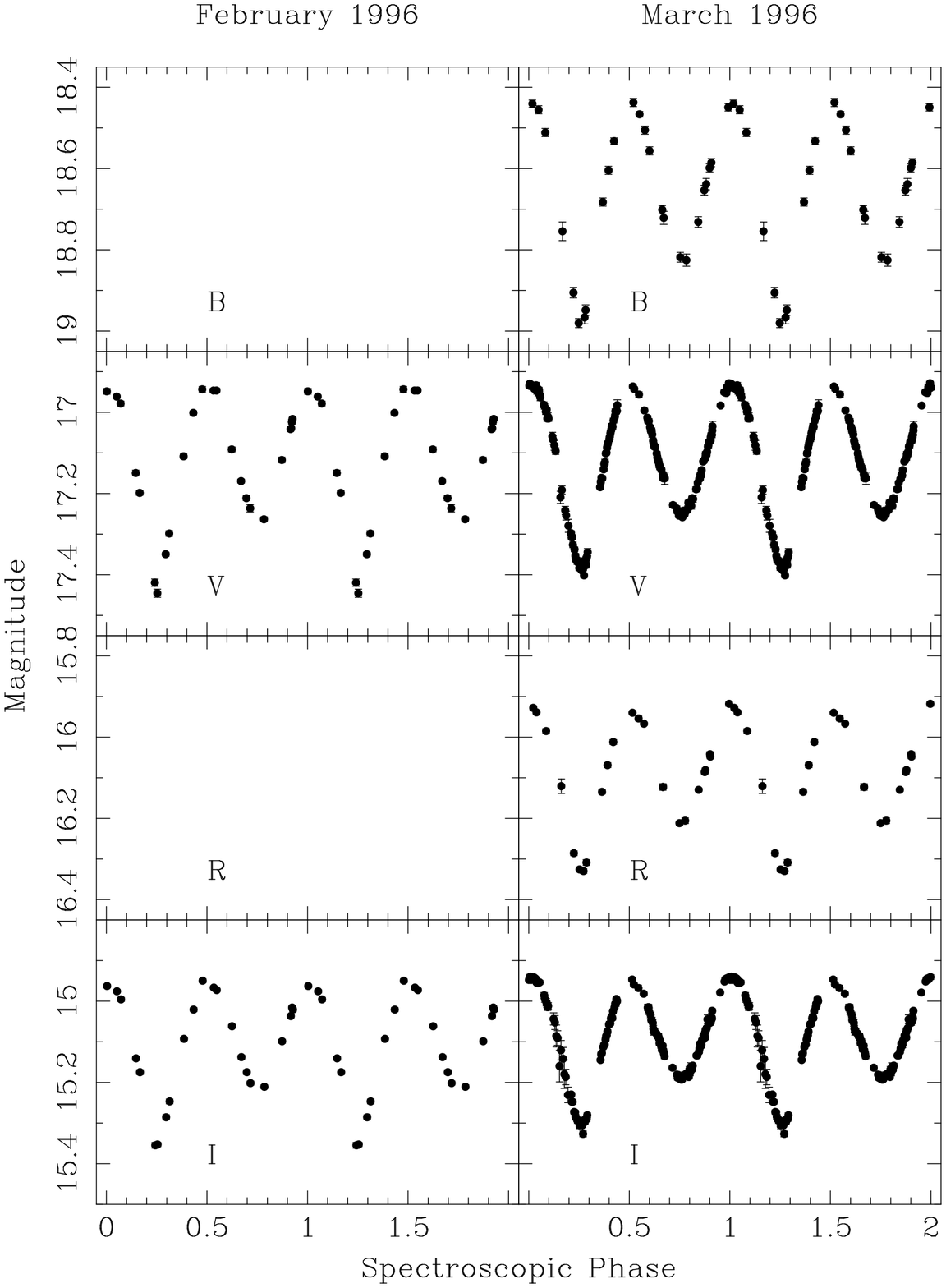} 
\caption{The light curves from February, 1996 (left panels) and
March, 1996 (right panels) folded on the spectroscopic
ephemeris (Table \protect\ref{sco1tab3}) are shown.  Each point
has been plotted for clarity.  Error bars are shown for all of
the points, but in many cases the size of the errors is smaller than
the symbols.}
\label{sco1fig3}
\end{figure*}

The light curves from February and March, 1996 folded on the spectroscopic
ephemeris are shown in Figure \ref{sco1fig3}.  The light curves are dominated
by ellipsoidal variations with maxima at the spectroscopic phases
0.0 and 0.5 (the quadrature phases) and minima of unequal depth at
the spectroscopic phases 0.25 and 0.75 (the conjunction phases).
During March, 1996 we observed three local extrema:  a deep minimum
on March 22, a maximum on March 24, and a shallow minimum on March 26.
We fit a parabola to the $V$ data from each of these three nights and
determined the times of the local extrema from the fits.  The times
are (HJD 2,450,160+) $4.835\pm 0.010$, $6.806\pm 0.010$, and
$8.766\pm 0.010$.  The spectroscopic phases of these three times are
$0.261\pm 0.008$, $0.013\pm 0.008$, and $0.760\pm 0.008$, very close to
their expected values of 0.25, 0.00, and 0.75.  Thus, the phasing of the
March, 1996 $V$ light curve is consistent with the spectroscopic ephemeris
given in Table \ref{sco1tab3}.

Our revised spectroscopic period can be compared with the period of
$2\fd 2616\pm 0\fd 0016$ found by 
van der Hooft et al.\ (1996)\markcite{vdf96} 
based on photometry from May to July, 1995.  Our period
agrees with theirs to within their errors.  However 
van der Hooft et al.\ (1996)\markcite{vdf96} 
then used spectroscopic fiducial points (including the one in
Bailyn et al.\ 1995b\markcite{bomr95b}) 
and derived the following ephemeris:
\begin{equation}
T_{\rm min}({\rm HJD})=
2\, 449\, 838.4279(30)+ 2.62032(50)N
\label{vander}
\end{equation}
This period differs from our period by
$2.5\sigma$.  Also, the phases of the three local extrema in our
$V$ light curve from March 1996 are by the
above ephemeris $0.568\pm 0.024$, $0.320\pm 0.024$, and $0.068\pm 0.024$,
significantly different from their expected values of 0.50, 0.25, and 0.00.
Bailyn et al.\ (1995b)\markcite{bomr95b} 
reported great difficulties in phasing up their
light curves from March and April, 1995---no single period could align
all of the optical minima with the radial velocities.  It appears that
the light curves from May to July, 1995 still suffer from phasing problems,
although to a much smaller extent than the earlier light curves from 1995.
The period derived solely from the radial velocities appears to be the most
reliable.

\begin{figure*}[t]
\plotone{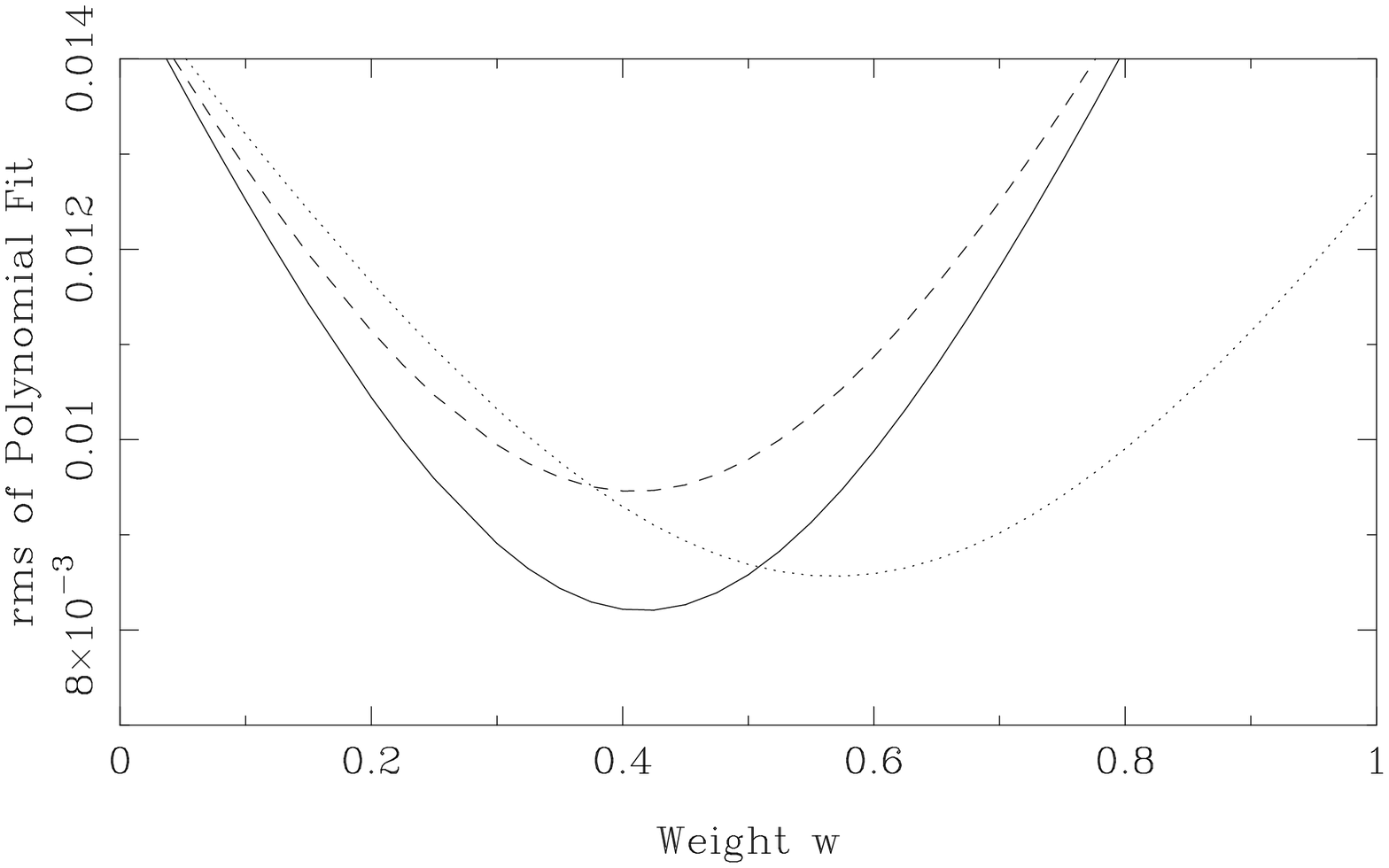}  
\caption{The rms residuals of a polynomial fit to the difference spectrum as
a function of the weight $w$ for three different template stars:
an F3IV star (solid line), an F3V star (dotted line), and a G0V star
(dashed line).  In each case, the minimum of the curve can be accurately
computed.
The restframe spectrum was from 1996.  See the text for more
details.}
\label{para}
\end{figure*}

The refined period of $P=2\fd 62157\pm 0\fd 00015$ represents a great
improvement in accuracy over the previous value given in 
Bailyn et al.\ (1995b)\markcite{bomr95b}.  
Using this new period, we can now compute accurately the
orbital phase of observations made in 1994 and late 1995.  For example,
Bailyn et al.\ (1995a\markcite{ba95a}) 
presented photometry from August 17, 1994
which showed evidence for an eclipse (the source got fainter and redder
then bluer and brighter over the 4.5 hours of observations). 
Only one such eclipse-like event was observed in 1994.  Using the refined
orbital period, we find that the spectroscopic phase of the 
time of minimum light (estimated to be at the heliocentric
Julian date
2,449,581.63) is 0.80.  
Curiously enough, the sharp optical minimum observed April 2, 1995
(see 
Bailyn et al.\ 1995b\markcite{bomr95b}) 
also has a spectroscopic phase of 0.80
(the phasing offsets of the light curves will be discussed more below).
The ASCA X-ray satellite observed GRO J1655-40 at the end of
August 23, 1994 
(Inoue et al.\ 1994\markcite{inisu94}).     
During most of this observation the X-ray
flux stayed relatively low.
Also, the X-ray spectrum from August 23 was quite different when compared
to subsequent observations, leading
to the speculation that an X-ray eclipse was observed (Inoue 1995, private
communication).  It turns out that the spectroscopic phases of
the August 23 ASCA observations are from 0.27 to 0.36---i.e.\ the star was
{\em behind} the 
compact object.  Thus some other mechanism must be invoked to
explain the X-ray light curve and spectrum from August 23, 1994.

\begin{figure*}[t]
\plotone{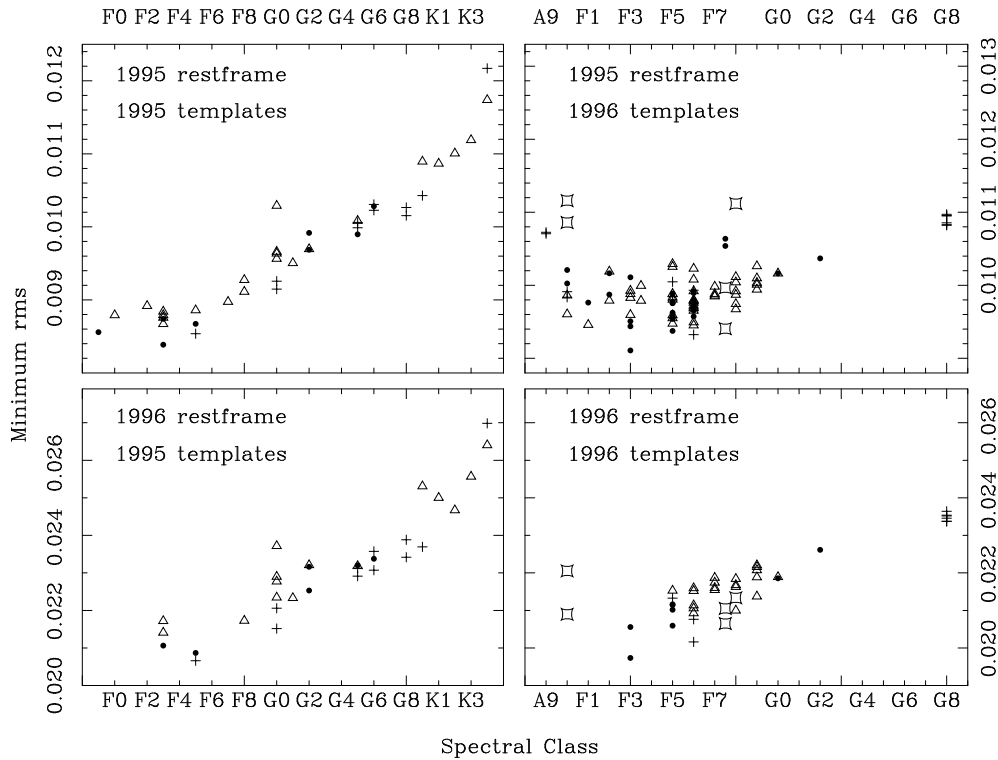}  
\caption{The minimum rms (see the text) is plotted as a function
of the spectral type of the template for the four different groups.
The plotting symbol indicates the luminosity class of the template:
open triangles for dwarfs, filled
circles for subgiants, crosses for giants, and  curved
squares for supergiants.  
Templates that had a weight $w>1$ are not shown (see the text).
In each case, the lowest {\em overall} rms values
occur for the spectral classes F3 to F5.  Note that many of the template
stars had two or more observations, so that the total number of template
stars used is less than the number of symbols shown.}
\label{plotMK}
\end{figure*}

\section{Spectral Classification}\label{secspec}

During the April-May, 1995 observing run, the spectra of 33 different
stars of spectral type F through K and luminosity classes V through
III were taken.  During the February, 1996 run, we obtained the
spectra of 45 different stars,
including nine 
Morgan-Keenan (1973)\markcite{mk73} 
spectral type standard stars
of class F.  We used the technique outlined in
Orosz et al.\ (1996)\markcite{obmr96}
to classify the 1995
and 1996  spectra of GRO J1655-40.
First, each continuum-normalized spectrum of GRO J1655-40 is
shifted to zero velocity and the lot of them are averaged together, creating
a ``restframe'' spectrum.  
To help minimize the effects of an occasional cosmic ray, a ``min-max''
rejection scheme was used where the lowest and highest values at each pixel
were rejected before the average was computed.
Then, the velocity shifts between the comparison
spectra and the restframe spectrum are removed.  Finally,
the steps involved in the comparison are
(1)~each continuum-normalized and shifted
comparison spectrum is scaled by a factor
$w$ (where $0\leq w\leq1$) and subtracted from the normalized 
restframe spectrum, (2)~a low order polynomial is fit to the difference
spectrum, (3)~the {\rm rms} residuals of the fit (computed over the same
region used to compute the cross-correlations) is recorded, and (4)~steps
(1) through (3) are repeated using different scaling factors
 until the minimum
{\rm rms} is found,
corresponding to the ``smoothest'' difference spectrum.
In finding the rms of the fits to the difference spectra, we were careful
to avoid interstellar absorption lines and regions corrupted by
bad columns on the CCDs.  Figure \ref{para} shows a plot of the rms 
of the fit to the difference spectrum as a function of the weight
$w$ for some representative cases.  In each case, the minimum of the curve
can be determined accurately.
  All of the above four steps are done on all of the comparison
spectra, and the comparison spectrum that has the
lowest overall {\rm rms} value is taken to be the spectrum whose
absorption lines most closely match those in the restframe spectrum.
The relative fluxes of the restframe spectrum and the difference spectrum
give an estimate of the fraction of the total light due to the accretion disk. 

\begin{figure*}
\plotone{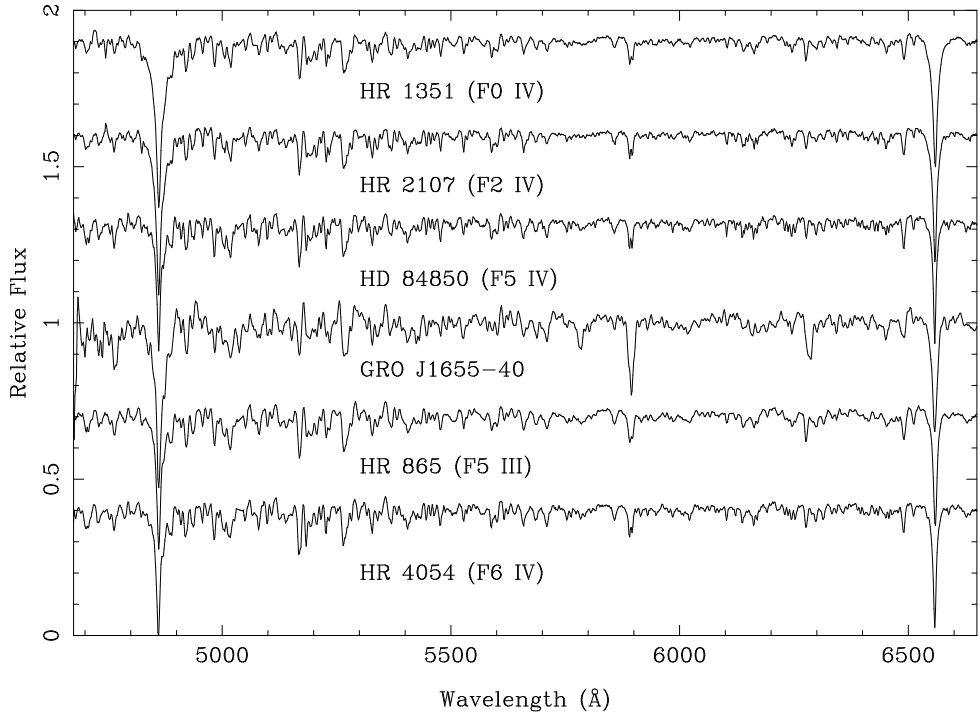}  
\caption{The
restframe spectrum of GRO J1655-40 from
February, 1996 (fourth from the top) 
and the spectra of several F subgiants and giants.  The anomalous strengths
of the absorption lines near 5900~\AA\ and 6300~\AA\  in the spectrum of
GRO J1655-40 are due to interstellar absorption.  Each spectrum has been
normalized to its continuum fit, and offsets have been applied for clarity.
}
\label{sco1fig2}
\end{figure*}

\begin{deluxetable}{lll}
\tablewidth{0pt}
\tablecaption{Input Model Parameters}
\tablehead{
parameter &
description &
comment}
\startdataT
$i$          & inclination &     free                        \nl
$Q$          & mass ratio ($M_1/M_2>1$)   &  free            \nl
$r_d$        & outer radius of the disk\tablenotemark{a} & free             \nl
$r_{\rm inner}$ & inner radius of the disk\tablenotemark{a} & fixed at 0.005\nl
$T_{\rm disk}$        & disk temperature at the outer edge & free                \nl
$\xi$       & power law exponent on disk &                       \nl
             & \quad temperature radial distribution & free          \nl
$\beta_{\rm rim}$ & flaring angle of the disk rim  &  free     \nl
$L_x$        & X-ray luminosity of the compact object & fixed at 0.0\nl
$W$          & X-ray albedo      &  fixed at 0.5          \nl
$T_{\rm eff}$  & polar temperature of the secondary &  fixed at 6500 K \nl
$\beta$      & gravity darkening exponent &   fixed at 0.25 \nl
$u(\lambda)$ & linearized limb darkening parameter & fixed\tablenotemark{b} \nl
$P$          & orbital period         &  fixed at $2\fd 62157$ \nl
$f(M)$       & mass function          &  fixed at $3.24\,M_{\sun}$ \nl
$\Omega$     & ratio of rotational angular velocity &    \nl
             & \quad to orbital angular velocity    & fixed at 1.0 \nl
$f$          & Roche lobe filling factor         & fixed at 1.0 
\enddata
\tablenotetext{a}{The disk radii are scaled to the effective Roche lobe
radius of the compact object.}
\tablenotetext{b}{Limb darkening parameters from 
Al-Namity (1978)\markcite{al78}
and
Wade \& Rucinski (1985)\markcite{wr85}.}
\label{sco1tab4}
\end{deluxetable}

We have two restframe spectra of GRO J1655-40, one each from 1995 and 1996
(for the 1995 restframe spectrum, we did not include the spectra from
May 2, 1995 because the star was partially eclipsed by the disk).
We also have two sets of template comparison spectra from the same times.
Every spectrum in each set of templates was compared against both restframe
spectra.  Since the 1995 spectra have a slightly higher spectral resolution 
than the 1996 spectra 
and since the 1996 restframe spectrum has a lower signal-to-noise ratio than
the 1995 restframe spectrum, a given template
spectrum will have a slightly higher
minimum rms value when compared against the 1996 restframe spectrum.  
We therefore grouped the comparisons into four different cases: the
1995 template spectra compared against the 1995 restframe spectrum, the
1995 template spectra compared against the 1996 restframe spectrum, etc.
The relative values of the minimum rms within each group can be used to
judge the best spectral match to the given restframe spectrum.
Figure \ref{plotMK} shows the minimum rms value plotted against the spectral
type of the template spectrum for the four different cases.   In all four 
cases, there is a clear trend: the best matches for the GRO J1655-40
restframe spectra are the F3-F5 stars.  Stars with spectral types later than
F6 or earlier than F3 provide much poorer matches to the restframe spectra.
Based on the values of the weight $w$ found for the best matches, we
conclude that the accretion disk contributed about 50\% of the flux
in $V$ during April and May, 1995, and less than 10\% of the flux in
$V$ during February, 1996.

This comparison technique is relatively
sensitive to the temperature class of the comparison star and
relatively insensitive to the luminosity class of the comparison
star.  We therefore cannot reliably
determine the luminosity class of the secondary
star from this technique alone.
However, in the case of GRO J1655-40, a main sequence
F star is much too small to fill the Roche lobe 
(Bailyn et al.\ 1995b\markcite{bomr95b}), 
so the secondary star must be somewhat evolved.
Also, in the case of the 1996 restframe spectrum, many of the early F dwarf
template stars
were excluded because a weight of $w>1$ was required to get the minimum
value of the rms.  Hence, we are left with mainly subgiants and giants.
So, based on these other pieces of evidence, we conclude the secondary
star in GRO J1655-40 is a subgiant, giving its full spectral classification
as F3-F5 IV.
Indeed, with the exception of the strong interstellar
features, it is difficult to tell the difference between the 1996 restframe
spectrum of GRO J1655-40 and the spectra of the nearby
F subgiants and giants
(see Figure \ref{sco1fig2}).

The 1995 spectra of GRO J1655-40 were examined in more detail to see if
the best fitting spectral type depends on the orbital phase.  There was
no indication of any change in the spectral type as a function of phase.
Further details can be found in 
Orosz (1996)\markcite{orosz96}.

\section{Light Curve Models}

\begin{deluxetable}{llll}
\tablewidth{0pt}
\tablecaption{Fits to the March, 1996 photometry}
\tablehead{
parameter &
model 1 &
model 2\tablenotemark{a} &
model 3  
}
\startdataT
$i$ (degrees)  &  $69.50 \pm 0.08$  &$74.7\pm 1.2$  & $69.54\pm 0.08$ \nl
$Q$            &  $2.99\pm 0.08$    &$3.3\pm 0.5$    & $3.50\pm 0.08$ \nl
$r_d$          &  $0.747\pm 0.010$  & $0.71\pm 0.01$   & $0.748\pm 0.011$ \nl
$T_{\rm disk}$ (K) & $4317\pm 75$   & $4525\pm 102$   & $4429\pm 75$ \nl
$\beta_{\rm rim}$ (degrees)& $2.23\pm 0.18$ & $1.54\pm 0.4$ &$2.91\pm 0.18$\nl
$\xi$          &  $-0.12\pm 0.01$   & $-0.22\pm 0.01$  &$-0.13\pm 0.01$ \nl
$L_x$          & 0.0 (fixed)        &0.0 (fixed)&  0.0 (fixed) \nl
$W$& 0.5 (fixed)        &0.5 (fixed)& 0.5 (fixed)  \nl
$T_{\rm eff}$ (K) &6500 (fixed) &6500 (fixed) &7000 (fixed) \nl
$\chi^2_{\nu}$ & 1.1551   &  3.2391       & 1.1816
\enddata
\tablenotetext{a}{No checking for eclipses done.}           
\label{sco1tab5}
\end{deluxetable}

The light curves shown in Figure \ref{sco1fig3} appear to be almost
completely ellipsoidal---there are two equal maxima per orbit and
two minima of unequal depth
per orbit. The minimum at the spectroscopic phase
0.25 (when the star is behind the compact object)
is deeper because the gravity darkening near the $L_1$ point
is greater and hence the star appears darker
(see 
Avni 1978\markcite{av78} 
and the Appendix).  
The symmetry and smoothness of the GRO J1655-40 light curves from 1996
are in strong contrast
to the light curves for the other black hole binaries
(e.g.\ 
McClintock \& Remillard 1986\markcite{mr86}; 
Wagner et al.\ 1992\markcite{wkhs92};
Chevalier \& Ilovaisky 1993\markcite{ci93}; 
Haswell et al.\ 1993\markcite{hrhsa93}; 
Haswell 1996\markcite{ha96};
Remillard et al.\ 1996b\markcite{romb96}; 
Orosz et al.\ 1996\markcite{obmr96}), 
where the light curves
are complicated by ``flickering'' about the mean light curve and by
large asymmetries in the light curve that slowly change with time
(e.g.\ 
Haswell 1996\markcite{ha96}).  

Because the light curves from February and March, 1996 are smooth and
symmetric and because the luminosity of the star dominates (i.e.\ the 
fraction of the light from the disk is small, see 
Section \ref{secspec}), we have a unique opportunity to model the
light curves and obtain a reliable constraint on the orbital inclination.
We have developed a detailed code based on the work of Avni 
(Avni \& Bahcall 1975\markcite{ab75}; 
Avni 1978\markcite{av78})
to model the light curves, which
is fully described in the Appendix.  The code uses full Roche geometry
to account for the distorted secondary, and light from a circular
accretion disk is included.  The code also handles mutual
eclipses by the star and the disk.  The effects of 
X-ray heating on the secondary star
are included as well.  There are several input model parameters
which are summarized in Table \ref{sco1tab4}.  See the Appendix for
detailed discussions of these parameters and their meaning.

For simplicity, we have fixed several of the input parameters at reasonable 
values.  For example,
we know the spectral type of the star (see the previous section), from which
we can get its effective temperature. 
The effective temperature of 6500 K is appropriate
for an  F5 IV star 
(Strai\v{z}ys \& Kuriliene 1981\markcite{sk81})
and we will adopt the effective temperature of
6500 K as the polar temperature.  
The gravity darkening
exponent is fixed at 0.25 since the star has a radiative envelope
(see the Appendix).  
The limb darkening coefficient
$u(\lambda)$ is better determined from model atmosphere 
computations, and we used values interpolated from tables
given by 
Wade \& Rucinski (1985)\markcite{wr85}.  
For fits to the 1996 light curves, we have fixed the
X-ray luminosity at 0, based on the ASCA
observation in early 1996 that
found $L_x\approx 2\times 10^{32}$~ergs~s$^{-1}$, 
almost three orders of magnitude lower
than the optical luminosity of the secondary star 
($L_{2}\approx 47\,L_{\sun}$,
see the next Section).  
In the case where $L_x=0$, the exact value of the
X-ray albedo $W$ is of course irrelevant.  For fits to the March 1995
light curves where $L_x\gg L_{2}$, we have adopted $W=0.5$
(see the Appendix),
although other values of the X-ray albedo $W$ have
been used (for example 
van der Hooft et al.\ (1996)\markcite{vdf96} 
used $W=0.4$).
Since there is mass transfer taking place, we assume the star fully
fills its Roche lobe and that it is in synchronous rotation.

For fits to the 1996 light curves, we are left with six free parameters.
The six free parameters in the model
are the inclination of the orbit ($i$), the mass ratio ($Q\equiv M_1/M_2$),
the outer radius of the disk in terms of the primary's effective
Roche lobe radius ($r_d$), the temperature at the outer edge of the
disk ($T_{\rm disk}$), the flaring angle of the rim of the disk
($\beta_{\rm rim}$, where the thickness of the disk at the outer edge is
given by $2r_d\tan\beta_{\rm rim}$), and the power law exponent on the
disk's radial temperature distribution ($\xi$, where the disk temperature
varies with radius as $T(r)=T_{\rm disk}(r/r_d)^\xi$).

\begin{figure*}[p]
\epsscale{1.7}
\plotone{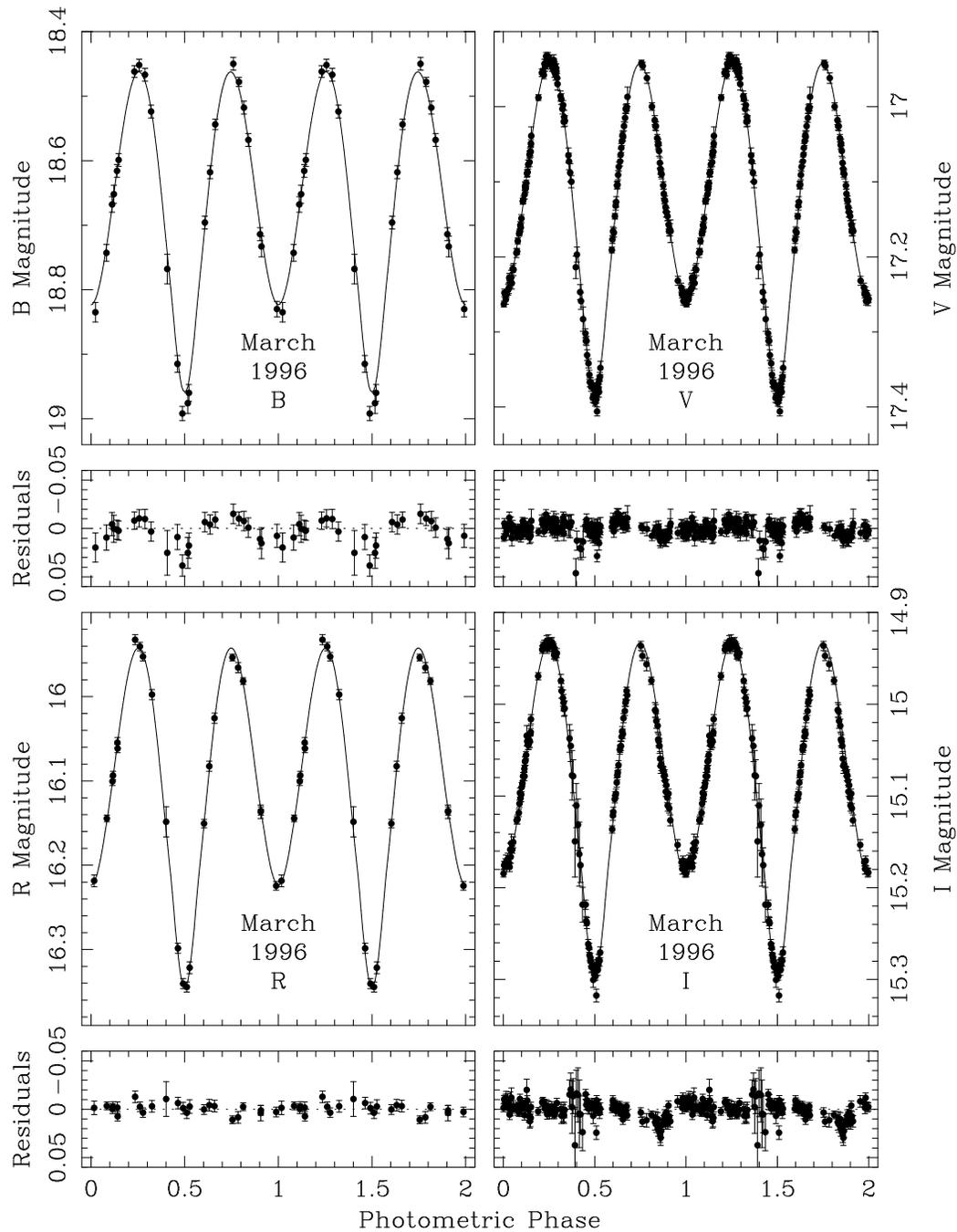}  
\caption{The model fits to the $B$, $V$, $R$, and $I$ light curves from
March, 1996 (large panels) and the residuals of the fits
(i.e.\ the data {\em minus} the model---small panels).
Table \protect\ref{sco1tab5} gives the model parameters under ``model 1.''
}
\label{mar96BtoI}
\end{figure*}

We first fit models to the $BVRI$ light curves from March, 1996, as the phase
coverage and spectral coverage from that time is the most complete.
We folded the data on the photometric phase
convention used in the code where the
photometric phase 0.0 corresponds to the time the secondary is directly
in front of the compact object (i.e.\ $T_0({\rm photo})
=T_0({\rm spect})+0.75P$).  As discussed in Section 3, the phases of
the three local extrema observed in March, 1996 are all late in
phase by $\approx 0.011$.  We found that the model fits were slightly better
after a phase shift of 0.011 was removed from the folded curves.  Since
there are far fewer points in the $B$ and $R$ filters, we gave each point
in $B$ and $R$ seven times more weight when computing the chi-square of the
fit.  This procedure resulted in equal weight being given to each filter.
Figure \ref{mar96BtoI} shows the fits and the residuals and Table
\ref{sco1tab5} lists the model parameters (under ``model 1'').  Note that
the data in all four filters were fit simultaneously.
The model fits the observed light curves from all four filters quite well---the
scatter in the residuals is typically less than 0.02 magnitudes.
The $1\sigma$ statistical errors for each parameter were estimated
using primarily a Monte Carlo ``bootstrap'' method 
(Press et al.\ 1992\markcite{ptvf92}---see 
the Appendix).
For this model, the accretion disk contributes about 5\% of the flux
at 5500~\AA, consistent with the measurement of $\leq10\%$ from late
February, 1996 (see the previous Section).

\begin{figure*}  
\plotone{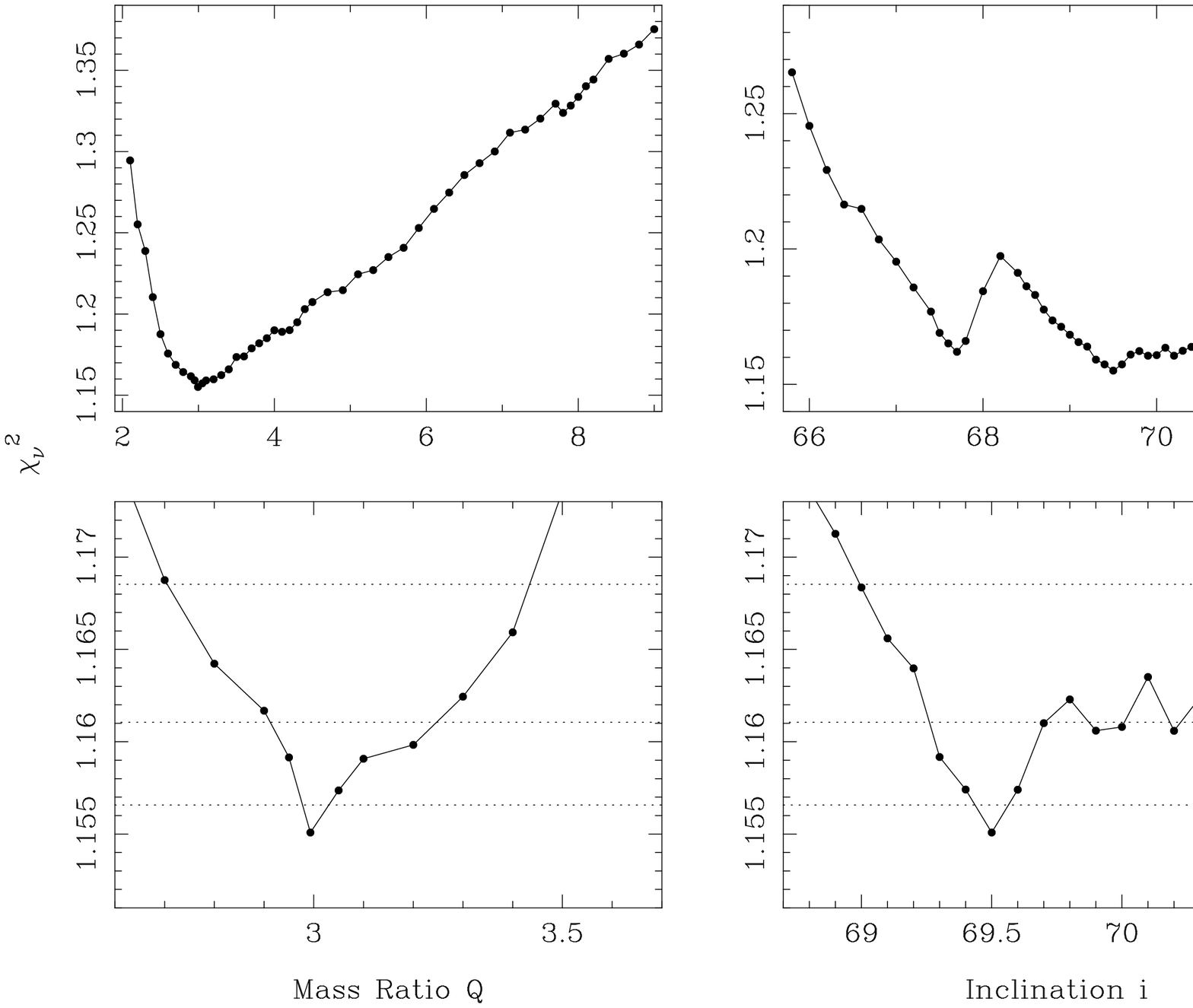}  
\caption{Upper left:
The value of the reduced chi-square
as a function of the mass ratio $Q$, where the value of $Q$ is fixed
at each input value and the other five parameters are adjusted until
the chi-square is minimized.  Lower left:  An expanded view of the
$\chi^2_{\nu}$ vs.\ $Q$ curve.  The dotted lines indicate the change in
the chi-square required for a 1, 2, and $3\sigma$ change in a single parameter
(i.e.\ $\Delta\chi^2=1$, 4,
and 9).  Upper and lower right: similar to the left
panels, but for fits with the inclination angle $i$ fixed at
various input values.}
\label{Qfix}
\end{figure*}

As a test of the accuracy of the $1\sigma$ errors derived from the bootstrap
analysis,
we performed fits the the March 1996 light curves
with 
the value of $Q$ at fixed at
several different input values.  The other five parameters were free
and adjusted to find the minimum reduced chi-square at each input value
of $Q$.  The left panels of Figure 
\ref{Qfix} shows the $\chi^2_{\nu}$~vs.~$Q$
curve.  The dashed lines shown in the lower left panel
of Figure \ref{Qfix} indicate the change in $\chi^2_{\nu}$ representing a 
1, 2, and $3\sigma$ change in a single parameter
(i.e.\ $\Delta\chi=1$, 3, and 9).  A change in $Q$
of 0.08 (the $1\sigma$ error derived from a Monte Carlo bootstrap analysis) 
in either direction from its value at the minimum causes the
$\chi^2_{\nu}$ to increase by approximately the amount indicated by
the lowest dashed line.   
The right panels of Figure \ref{Qfix} show the equivalent computation done with
the value of $i$ fixed at several different values.  Like before, a $1\sigma$
change in the parameter $i$ (0.08) from the best value causes the value of 
$\chi^2_{\nu}$ to increase by approximately the amount indicated
the lowest dashed line in the lower right panel.
The $\chi^2_{\nu}$~v.s.~$i$ curve has a secondary minimum at
$i=67.7\deg$ and $\chi^2_{\nu}=1.1620$, which is greater than the
$2\sigma$ deviation from the primary minimum at $i=69.5\deg$ and
$\chi^2_{\nu}=1.1551$.  While
we are confident that the errors derived from the bootstrap analysis are
a reasonable representation of the true internal statistical errors,
there may be systematic errors due to physical effects not
included in the model.  Given how well the model fits the data, we
suspect that such systematic errors are likely to be small.
We are also confident that we have found the global minimum of $\chi^2_{\nu}$
since we have searched a large amount of parameter space.

For each model with $Q$ fixed at a given input value, there exists a value of
the inclination $i$ found from the best fit.  For the whole range of $Q$ shown 
in the upper left panel of Figure \ref{Qfix}, the corresponding best
values of $i$ range from $68.40\le i\le 71.1\deg$.  Likewise,
for the models with the inclination angle $i$ fixed at several values
(the upper right panel of Figure \ref{Qfix}),
the corresponding best values of the mass ratio $Q$ are in the range
$2.52\le Q\le 4.23$.  Thus, the key geometrical parameters $i$ and $Q$
vary little and seem to be reasonably well determined.  Based on the
behavior of $\chi^2$ near the minimum, we adopt the $1\sigma$ and $3\sigma$
ranges of the inclination $i$:
\begin{eqnarray}
       69.42\deg \le & i & \le 69.58\deg  \quad (1\sigma)  \nonumber  \\
       69.0\deg  \le & i & \le 70.6\deg  \quad  (3\sigma)  \nonumber
\end{eqnarray}
and the mass ratio $Q$:
\begin{eqnarray}
       2.91 \le & Q & \le 3.07  \quad (1\sigma)  \nonumber  \\
       2.60 \le & Q & \le 3.45  \quad  (3\sigma).  \nonumber
\end{eqnarray}

\begin{figure*}
\epsscale{1.7}
\plotone{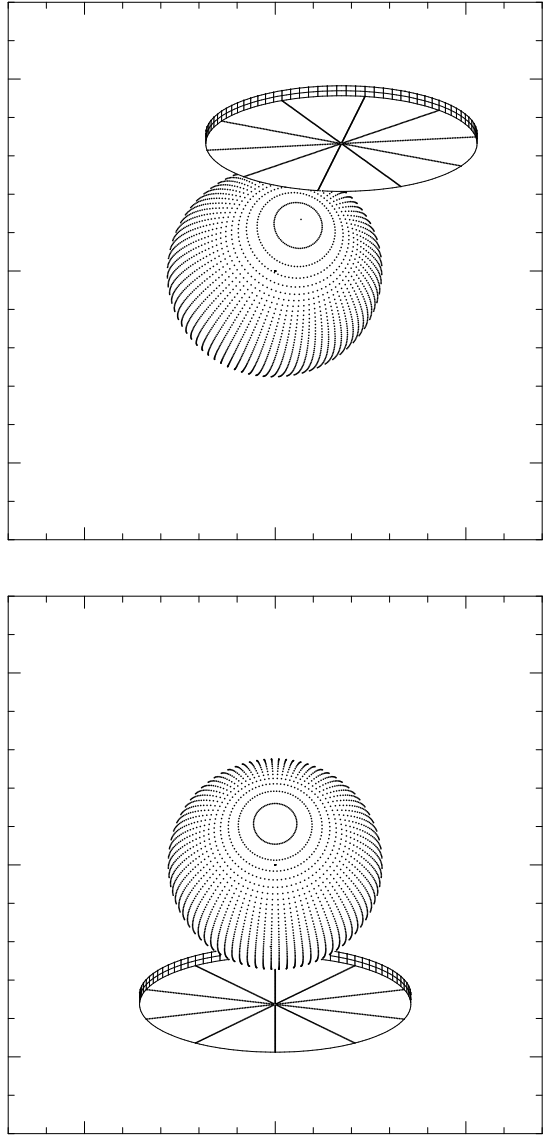} 
\caption{The system as it appears projected onto the plane of
the sky for
the model parameters given in Table \protect\ref{sco1tab5}, model 1.
The photometric phase is $0\deg$ for the bottom panel and $170\deg$
for the top panel.}
\label{geofig}
\end{figure*}

The geometry of the system is shown schematically for two different
phases in Figure \ref{geofig}.  Note that the compact object is
{\em not} eclipsed by the secondary star.  We would therefore {\em not}
expect to see X-ray eclipses (assuming the X-rays are emitted from a relatively
small region centered on the compact object), and in fact no
X-ray eclipses have been seen (e.g.\ 
Harmon et al. 1995a\markcite{ha95a}).  
The star does block part of the disk (and vice-versa) at certain phases,
so we expect to
still see grazing optical eclipses, the depths of which
depend on the size of the disk and the relative brightnesses of each
component.  In order to assess how important the grazing eclipses
are in determining the final overall shapes of the light curves,
we performed a model fit to the $BVRI$ data from March, 1996
where the eclipse checking routines were turned off.  In this case the
model light curve is just the normal ellipsoidal light curve from
a Roche lobe filling star plus a constant flux from the accretion disk
(where the amount of disk light relative to the amount of star light
in general depends on the filter bandpass).
Table \ref{sco1tab5} gives the values of the parameters (under
``model 2'').  The fit is poor, as one can deduce from the relatively large
value of the reduced chi-square ($\chi^2_{\nu}=3.239$,
compared to $\chi^2_{\nu}=1.1551$ for model 1). The fits are
poor when eclipses are not accounted for in the computations
for two reasons:
(i)~the difference in the depths of the two minima in the model light
curve are too
small,  and (ii)~the dependence of the light curve amplitude with
color is not properly accounted for by
the model
(the amplitude of the observed light curve
is largest in $B$ and smallest in $I$, see Figure \ref{sco1fig3}).

Our numerical experiments also show that the final fitted parameters do
not depend strongly on the assumed secondary star parameters.
For example, in the case of
a Roche lobe-filling
star, the temperature will not be  constant over the surface 
(see Equation (\ref{neweq8}) of the Appendix).  In
particular, the side of the star near the $L_1$ point should be the
coolest.  The  temperature of
6500 K corresponding to the observed
spectral type therefore represents some kind of flux-weighted average
of the  temperatures over the
surface, and technically not the star's polar temperature.
However, our spectral coverage was not uniform in phase and
spectra used to determine the spectral type mainly came from
phases near the quadrature phases.  The temperatures corresponding to
the brightest parts of the projected stellar disk at these phases are
within  $\approx 200$ K of the polar temperature.  Also, we could not
detect any  change in the spectral type of the individual observations
as a function of phase (this includes the data from May 2, 1995 when the
star was behind the disk and the $L_1$ point was in full view---see Orosz
(1996)), which suggests other factors are at work when the observed spectrum
at a given phase is produced.
Thus, we believe we are not making a large error when adopting
the temperature derived from the observed spectral type as the polar 
temperature.  In this same vein,
we fit a model to the March, 1996 data where the polar
temperature of the secondary star was 7000~K, appropriate for
an F2 IV star 
(Strai\v{z}ys \& Kuriliene 1981\markcite{sk81}).  
The results for the
geometrical parameters, given
in Table \ref{sco1tab5} under ``model 3'', are quite similar to the results
using $T_{\rm eff}=6500$~K (model 1).  Finally, we compared the linearized
limb darkening parameters $u(\lambda)$ interpolated from tables given by
Al-Naimiy (1978)\markcite{al78}
and 
Wade \& Rucinski (1985)\markcite{wr85}.  
The shapes of the
$u(\lambda)$ curves were very similar---the only large differences
occurred for wavelengths blueward of the $B$ band.  As a result, the fits
to the $BVRI$ light curves of GRO J1655-40 were similar when
the limb darkening coefficients were used from either source.  

\begin{figure*}
\plotone{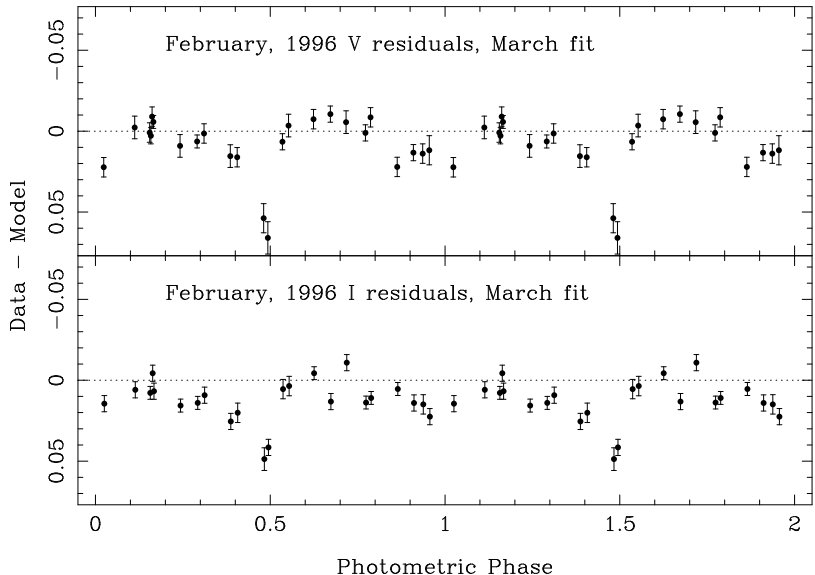} 
\caption{The two panels show the February, 1996 data {\em minus}
the March, 1996 model (Table \protect\ref{sco1tab5}, model 1).  
}
\label{febres}
\end{figure*}

The stellar evolution models (see the Discussion) indicate the secondary has 
a radiative envelope.  We have therefore adopted a gravity darkening 
exponent of $\beta=0.25$.  We tried a model fit to the light curves
with the gravity darkening exponent set to $\beta=0.08$, the value
appropriate for stars with convective envelopes (Lucy 1967).  The model
fits were much worse in this case (we could not find a fit with
$\chi^2_{\nu}<2$).  When $\beta=0.08$, the temperature contrast over the
star is smaller (see Equation (\ref{neweq8}) of the Appendix),
leading
directly to a smaller brightness contrast, especially between the side of
the star near the $L_1$ point and the opposite side. 
Since the brightness contrast between the side facing
the compact object and the opposite side is greatly reduced,
the inclination needs to be higher
to produce the same observed
amplitude of the light curves.   However,
as the inclination grows closer to $90\deg$, the eclipses become deeper and the
fine balance between the primary and secondary eclipse depths is disturbed,
making the {\em simultaneous} fits to all four colors much more difficult.
Thus, based both on the stellar evolution models and the quality of the fits,
we conclude that the gravity darkening exponent of $\beta=0.25$,
appropriate for a star with a radiative envelope, is the correct choice.

As one can see from Figure \ref{sco1fig3}, the shapes
of the February, 1996 light curves are almost identical to the shapes
of the light curves from March, 1996.  The only noticeable
difference between the two
curves occurs near 
the minimum at the photometric
phase 0.5 (which is the spectroscopic phase 0.25): that minimum
in the February light curves is slightly deeper
than that minimum in the March light curves. 
Not surprisingly, models that fit the March, 1996 data are consistent
with the February, 1996 data (Figure \ref{febres}):
the residuals in the $V$ band have a mean of 0.009 and a standard
deviation of 0.019 and the residuals in the $I$ band have a mean of 0.013
and a standard deviation of 0.013.
The only deviant points in the two filters are those
near the photometric phase 0.5
which are about 0.05 magnitudes
{\em fainter} than the model that fits the data from March.  

The light curves of GRO J1655-40 from March and April, 1995 (see 
Bailyn et al.\ 1995b\markcite{bomr95b}) 
look quite different from the light curves from 1996 (Figure
\ref{sco1fig3}).  First of all, the source was about 0.7 to 0.8
magnitudes brighter in $V$ during early 1995 than it was during early
1996, probably a direct result of the fact that there was still considerable
hard X-ray activity taking place during that time 
(e.g.\ 
Wilson et al.\ 1995\markcite{whzpf95}).
In 1995, the minimum near the spectroscopic phase
0.75 (when the secondary star is in front of the compact object)
is the deeper of the two, opposite of the situation in the 1996 light curves.
There are large asymmetries in the
1995 light curves, and the phases of the minima of the three light
curves shown in 
Bailyn et al.\ (1995b)\markcite{bomr95b} 
do not all line up properly
with the spectroscopic phase (the difference in phase is as
large as 0.05).  
Similar (but smaller) phase 
shifts in the optical light curve minima
have been seen in some low mass X-ray binaries
(for example X1822-371, 
Hellier \& Mason 1989)\markcite{hm89}.
The phase shifts of the minima are
presumably caused by large distortions in the accretion 
disk.  
Hellier \& Mason (1989)\markcite{hm89} 
invoked a thick accretion disk
where the thickness of the rim depended strongly on the azimuth angle
to explain the X-ray and optical light curves of X1822-371.  It is quite
clear that our simple model where the disk is a flattened 
azimuthally symmetric cylinder
will not be able to produce a model light curve whose minima are shifted
from their expected phases.  

\begin{figure*}
\plotone{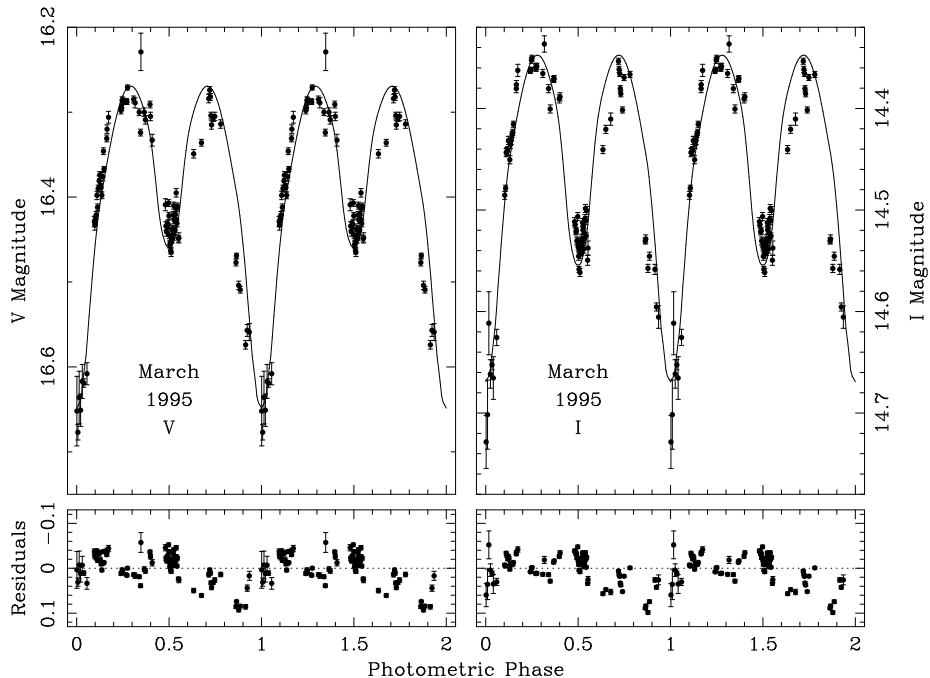}   
\caption{The models fits to the $V$ and $I$ light curves from March,
1995 (top panels) and the residuals of the fits
(i.e.\ the data {\em minus} the model---bottom panels).
Table \protect\ref{sco1tab6} gives the model parameters.
Note the scale changes in the plots compared to
Figure \protect\ref{mar96BtoI}.  These data were published previously in 
Bailyn et al.\ 1995b.}
\label{mar95VandI}
\end{figure*}

Out of the three light curves from 1995 presented in 
Bailyn et al.\ (1995b)\markcite{bomr95b}, 
the $V$ and $I$ light curves from March 18-25, 1995 are the
most symmetric.  In addition, the minima of these light curves line up
approximately at their expected phases when the data are folded on the revised
spectroscopic ephemeris presented in Table \ref{sco1tab3}
(whereas the minima of the later light curves do not).  
We attempted, a model fit to the March 18-25, 1995
data to see if the overall shape of those two light curves
can be explained by our model
(as discussed above we will not be able to model {\em all}
of the 1995 light curves because of the large phase shifts of the minima
of the other light curves).  
For fits to the March, 1995 light curves, 
we must take into account  X-ray heating of the
secondary star.  The part of the secondary star facing the compact object
will be hotter and hence brighter than it would otherwise be in the absence
of any external heating.
We set the X-ray albedo at $W=0.5$ and let the X-ray
luminosity of the compact object $L_x$ be a free parameter.
We fixed the inclination at $i=69.50\deg$ and the mass ratio at $Q=2.99$,
the values found from the model fit to the March, 1996 light curves.
The results are listed in Table \ref{sco1tab6}, and Figure
\ref{mar95VandI} 
shows the fits.  

\begin{deluxetable}{ll}
\tablewidth{0pt}
\tablecaption{Fits to the March, 1995 photometry}
\tablehead{
parameter &
{\begin{tabular}{c} March, 1995 \\  model 4 \end{tabular}} 
}
\startdata
$i$ (degrees)  &  $69.50$ (fixed)         \nl 
$Q$            &  $2.99$ (fixed)          \nl 
$r_d$          &  $0.95\pm 0.01$          \nl 
$T_{\rm disk}$ (K) & $6694\pm 100$         \nl
$\beta_{\rm rim}$ (degrees)& $12.6\pm 0.1$\nl 
$\xi$          &  $-0.18\pm 0.01$         \nl 
$L_x$ (ergs~s$^{-1}$) & $3.72\times 10^{36}$ \nl
$W$            & 0.5 (fixed)              \nl 
$T_{\rm eff}$ (K)  &  6500 (fixed)        \nl 
$\chi^2_{\nu}$ & 66.0                     \nl 
\enddata
\label{sco1tab6}
\end{deluxetable}

Considering the simplicity of the model,
the overall shapes of the $V$ and $I$ light curves are fit
reasonably well.  There is considerable ``flickering'' about the mean
light curve with deviations larger than the observational errors, 
resulting in the large value of the
reduced chi-square.
Note the difference
in the disk parameters between the fits to the March, 1995 and March,
1996 data:  the disk is much larger and hotter in 1995. 
It is interesting to note that the disk and the star contribute roughly
equal amounts of flux in the $V$ band for the model that fits the
March, 1995 data, which is in good agreement with the value of the disk
fraction estimated
from the spectra from April and May, 1995
(which were {\em not} simultaneous with the photometry).  
The observed X-ray luminosity, as determined from the BATSE daily
averages, varied by about a factor of four between March 18 and March 25,
1995, with $5.7\times 10^{36}\lesssim L_x \lesssim 2.3\times
10^{37}$~erg~s$^{-1}$ (S. N. Zhang, private communication).
Considering the simplistic way the X-ray heating is handled in the model,
the fitted value of
$L_x=3.7\times 10^{36}$~ergs~s$^{-1}$ 
is remarkably close to the observed range.
A slightly lower value of the X-ray albedo $W$ (like the value of
$W=0.4$ used by 
van der Hooft et al.\ 1996\markcite{vdf96}) 
would require a larger value
of $L_x$ to produce the same heating effect
(e.g.\ see Equation (\ref{neweq29}) of the Appendix), so we do not consider the
slight difference between the observed X-ray luminosity and the fitted
luminosity to be a problem.
Finally, we note that the depth and shape of the
minimum at the photometric phase 0.5 is sensitive to the amount of
X-ray heating.  Given that the observed hard X-ray flux from the source
varied by about a factor of four during March 18-25, 1995, 
we should not be surprised to see
a large amount of scatter in the optical light curves around phase 0.5.
In summary,
although we cannot fit every detail in the March, 1995 light curves,
we conclude that the basic overall shapes of the light curves
(i.e.\ the relative depths of the minima) can be explained by the effects
of X-ray heating on the secondary star.
To explain the phase-shifted minima of the other 1995 light curves,
we would probably have to invoke an accretion disk whose shape and
brightness profile are grossly distorted---something which is beyond the
scope of our current model.

The code also computes radial velocities.  For a Roche-lobe filling star,
the center of light does not precisely follow the center of mass of the star,
resulting in a slightly distorted radial velocity curve
(for example, see
Kopal 1959 \markcite{ko59} or
Wilson \& Sofia 1976\markcite{ws76}).
Partial eclipses of the star can result in an asymmetric contribution to the
star's rotational velocity to the overall observed radial velocity, leading
to a distorted radial velocity curve near the conjunction phases.  On
May 2, 1995, the secondary star was observed spectroscopically by
Bailyn et al.\ (1995b)\markcite{bomr95b}
as it moved through the spectroscopic phase 0.25 (i.e.\ as it moved directly
behind the disk).  The radial velocities from that night 
(Figure \ref{sco1fig1}) deviate systematically from the sine curve:  just
before the phase 0.25, the radial velocities are larger than expected and
just after the phase 0.25 the radial velocities are smaller than expected.
Stated another way, the residuals of the sine curve fit to the radial 
velocities (in the sense of ``data-model'') are positive just before phase
0.25 and negative just after phase 0.25.  As we discussed above, the accretion
disk must have had a complex and variable shape to account for some of the
features of the 1995 light curves.  Modelling the radial
velocity curve of GRO J1655-40 is made more difficult by the lack of 
simultaneous photometry.  Nevertheless,
the model shown in Figure
\ref{mar95VandI} does produce radial velocity residuals with the correct shape
near phase 0.25 (positive just before phase 0.25 and negative just after),
but the amplitude of the model residuals was only about $15$~km~s$^{-1}$, 
whereas the observed residuals have a much larger amplitude of $\approx
90$~km~s$^{-1}$.  Indeed, we could not explain the large amplitude
of the radial velocity residuals from May 2, 1995.  Further details of
the radial velocity curve models can be found in 
Orosz (1996)\markcite{orosz96}.

\section{Discussion}

\begin{deluxetable}{ccccc}
\tablewidth{0pt}
\tablecaption{Component masses for GRO J1655-40}
\tablehead{
{\begin{tabular}{c} confidence \\  limit \end{tabular}} &
{\begin{tabular}{c} $i$ \\  (degrees) \end{tabular}} &
$Q$                                                   &
{\begin{tabular}{c} $M_1$ \\  ($M_{\sun}$) \end{tabular}} &
{\begin{tabular}{c} $M_2$ \\  ($M_{\sun}$) \end{tabular}} 
}
\startdataT
$1\sigma$    &  69.42-69.58  &  2.91-3.07   &  6.80-7.24  &  2.20-2.46 \nl
$3\sigma$    &  69.00-70.60  &  2.60-3.45   &  6.42-7.63  &  1.86-2.90 \nl
\enddata
\label{masses}
\end{deluxetable}

The precision of our results for $i$ and $Q$ allow us to determine
the masses of the components to unprecedented accurately for a black
hole system
(see Table \ref{masses}).  
Indeed the formal precision of our results is so high that we
expect the true uncertainty of our results comes from physical effects
not included in the models.  However, given the good fit between our models
and the data, we expect that the systematic errors are also quite small.
In the following we discuss the implications of our results for the
nature of the secondary star, the radio jet, the evolutionary history of
the system, and the outburst mechanism.   We will use the parameters
listed in Table \ref{masses} for the system throughout the following 
discussion.

\subsection{Luminosity and Nature of the Secondary Star}

\begin{figure*}
\plotone{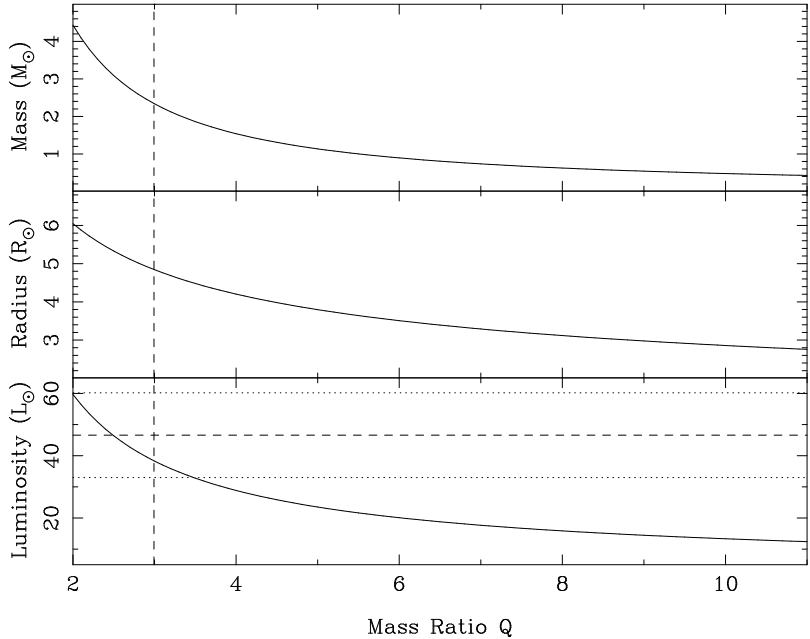} 
\caption{The mass (top), radius (middle), and the luminosity (bottom)
of the secondary star as a function of the mass ratio $Q$, assuming
$T_{\rm eff}=6500$~K,
$i=69.5\deg$, $f(M)=3.24\,M_{\sun}$, and $P=2\fd 62157$.
The dashed and dotted lines in the lower panel
indicate the observed luminosity of the secondary star in GRO J1655-40
and its $1\sigma$ error.}
\label{secparm}
\end{figure*}

The mass ratio of $Q=2.99\pm 0.08$ we found for GRO J1655-40 is the smallest
among the seven black hole systems
with low mass secondaries.  Much larger values of $Q$ have been found for the
three systems where the mass ratio has been directly computed via the 
measurement of the rotational broadening of the secondary star's absorption
lines: $Q=14.9$ for both A0620-00 
(Marsh, Robinson, \& Wood 1994\markcite{mrw94}) 
and V404 Cyg 
(Casares 1995\markcite{ca95}), 
and $Q=23.8$ for GS 2000+25 
(Harlaftis, Horne, \& Filippenko 1996\markcite{hhf96}).  
There are hints 
that the mass ratio
may be extreme in XN Mus91 
(Orosz et al.\ 1994)\markcite{obrmf94}, 
and GRO J0422+32
(Filippenko, Matheson, \& Ho 1995\markcite{fmh95}) 
as well.  In GRO J1655-40,   
we have a way to independently check the value of the mass ratio by
comparing the observed luminosity of the secondary
star to the luminosity inferred from the model fits. 
To compute the observed luminosity in $V$, we adopt a mean $V$ magnitude
of $\bar{V}=17.12$, which is the mean of the model fit to the March,
1996 data, a distance of $d=3.2\pm 0.2$~kpc (which is tightly constrained
by the kinematics of the radio jets---see
Hjellming \& Rupen 1995\markcite{hr95}), 
a disk fraction in $V$ of $5\%\pm 2\%$ (Section 4), and
a color excess of $E(B-V)=1.3\pm 0.1$ 
(Horne et al.\ 1996\markcite{ho96}; Horne
private communication 1996).  Assuming a visual extinction of
$A_v=3.1E(B-V)$, the observed luminosity in $V$
of the GRO J1655-40 secondary
is $L_{\rm obs}=46.6\pm 13.6\,L_{\sun}$.

The luminosity of the secondary star inferred from the models 
is easily derived from some simple relations.
Given the measured masses and orbital period of the system, we can use
Kepler's Third Law to determine the semi-major axis.  
Eggleton's (1983)\markcite{egg83}
expression for the effective radius of the Roche lobe then determines
the size of the secondary, and the temperature inferred from the spectral
type and the Stefan-Boltzmann Law determines the luminosity.
Figure \ref{secparm}
shows the computed mass, radius, and luminosity of the secondary
star as a function of $Q$, assuming an inclination of $i=69.5\deg$,
an effective temperature of $T_{\rm eff}=6500$~K, and an orbital period
of $2\fd 62157$.  The vertical dashed line in the three panels indicates
the value of $Q$ from the best model fit, and the horizontal dashed and
dotted lines in the lower panel indicate the observed luminosity of
the GRO J1655-40 secondary and its $1\sigma$ error.  Evidently, models with
$Q\lesssim 3.5$ are needed to produce a secondary star with a luminosity
large enough to match the observed luminosity of the GRO J1655-40 secondary.
Models with larger mass ratios (i.e.\ $Q\gtrsim 5$) not only
produce under-luminous secondary stars, but also provide fits
to the data with relatively large $\chi^2_{\nu}$ values (Figure 
\ref{Qfix}).  The agreement between the luminosity of the
secondary calculated from the orbital parameters and that
inferred from the known distance and reddening of the source provides a 
satisfying consistency check for our models.

\begin{figure*}
\plotone{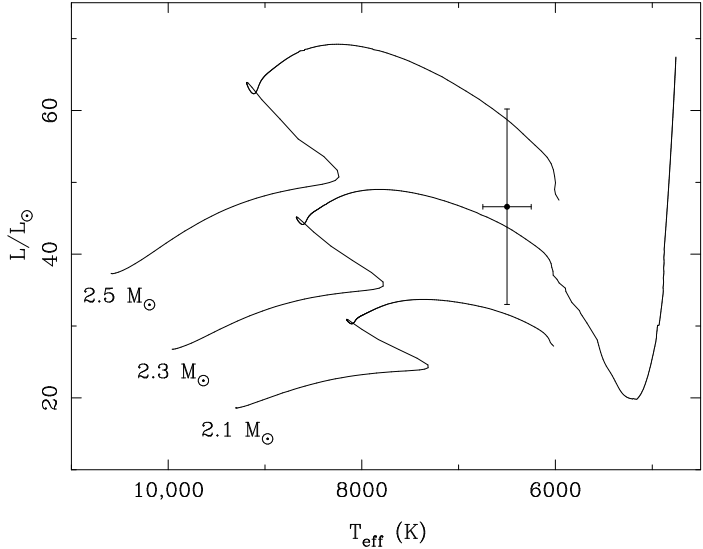} 
\caption{A Hertzsprung-Russell diagram showing the position of the secondary
star of
GRO J1655-40.  Evolutionary tracks are shown
for a $2.1\,M_{\sun}$
star (lower curve), 
a $2.3\,M_{\sun}$ star (middle curve), and a $2.5\,M_{\sun}$
star
(upper curve), assuming a
solar metallicity.  In each case, core hydrogen exhaustion
takes place just after the tiny loop.  The computations
of the tracks for the $2.1\,M_{\sun}$ star and $2.5\,M_{\sun}$ star
were stopped short of the giant branch.
}
\label{track}
\end{figure*}

The observed luminosity of the GRO J1655-40
secondary star is a fairly well-constrained quantity, owing to the fact
that the distance is well determined from the kinematics of the
radio jets 
(Hjellming \& Rupen 1995)\markcite{hr95} 
and the amount of interstellar extinction is well
determined from high quality UV spectra obtained with the {\em
Hubble Space Telescope} 
(Horne et al.\ 1996\markcite{ho96}).  
The observed
spectral type of the star provides a fairly good measure of its
effective temperature 
(e.g.\ 
Strai\v{z}ys \& Kuriliene 1981\markcite{sk81}), 
and 
we adopt $T_{\rm eff}=6500\pm 250$~K.
We therefore
can accurately
place the secondary star of GRO J1655-40 on a Hertzsprung-Russell
diagram (Figure \ref{track}).  Using the Yale Stellar Evolution Code
(Guenther et al.\ 1992\markcite{gdkp92}) 
with updated opacities
(Iglesias \& Rogers 1996)\markcite{ir96}, 
we computed the evolutionary
tracks of a $2.1\,M_{\sun}$ star, a $2.3\,M_{\sun}$ star, and
a $2.5\,M_{\sun}$ star, assuming a solar metallicity, a hydrogen abundance of
$71\%$, and a mixing length coefficient of 1.7, all typical values for
population I stars.
These evolutionary
tracks are also shown in Figure \ref{track}.  The secondary star 
falls very near
the track of the $2.3\,M_{\sun}$ star, roughly halfway
between the main sequence and the giant branch.  This provides yet another
consistency check for our model.
It is interesting to note
the radius of the $2.3\,M_{\sun}$
star near the position of the secondary:  690 Myr
after
the zero-age main sequence (ZAMS)
the star has a radius of $4.85\,R_{\sun}$
(and a temperature of $T_{\rm eff}=6813$~K and
a luminosity of $L_2=45.8\,L_{\sun}$);
692 Myr after
the ZAMS, the star has a radius of $5.21\,R_{\sun}$
(and $T_{\rm eff}=6500$~K and
$L_2=43.8\,L_{\sun}$).
For $Q=2.99\pm 0.08$ and $i=69.50\pm 0.08\deg$, the effective radius
of the secondary star is $4.85\pm 0.08\,R_{\sun}$ (when given the values
of the orbital period and mass function listed in Table \ref{sco1tab2}).
Thus, the observed luminosity, the observed
temperature, {\em and} the inferred
radius of the GRO J1655-40 secondary star are all consistent with a
$\approx 2.3\,M_{\sun}$ {\em single} star that has evolved 
$\approx 690\times 10^6$ years past the ZAMS.

\subsection{Inclination of the Radio Jet}

Hjellming \& Rupen (1995)\markcite{hr95} 
showed that the relativistic radio jets GRO J1655-40
possessed during late 1994 were inclined $85\pm 2\deg$ to the line of sight.
In addition, the jets were apparently rotating about the jet axis at an
angle of $2\deg$ and with a period of $3\fd 0\pm0\fd 2$.
Our light curve models show that the orbital plane is inclined $69.5\deg$
to the line of sight.  Hence, the jet axis was tilted by $15.5\deg$ from the
normal of the orbital plane.  {\em If} the jet axis was perpendicular to
the orbital plane, the $2\deg$ offset might be explained by a warped
accretion disk near the compact object.  In this case, the roughly three
day periodicity would be closely related to the orbital period of the binary
since the orientation of the warp would depend on the orbital phase.
However, since the jet axis is {\em not} perpendicular to the orbital plane,
it  appears difficult to account for both the $15.5\deg$
tilt in the jet axis {\em and} the $2\deg$ rotation about the axis by
invoking a simple warp in the inner accretion disk.  
If the inner disk were warped by $15.5\deg$ and the axis of the jet
were perpendicular to the warp, we might expect to observe a large
precession of jets because
the orientation of the
warp depends on the orbital phase.

Alternatively,
the $15.5\deg$
tilt of the jet axis may be a measure of the inclination of the spin
axis of the black hole with respect to the orbital plane.  
In this hypothesis the jet would need to be collimated in some way
by the Kerr geometry of the black hole itself.
The ``wiggles'' observed in the jets might be due to
the low mass ratio 
of the system
which results in a considerable displacement of the black hole 
from the
center of mass of the binary system: $4.19\pm 0.19\,R_{\sun}$.  Thus, if
the jet axis is aligned with the black hole, the base of the jets would
orbit about the center of mass with a maximum displacement of 
$\approx 8.4\,R_{\sun}$.  A highly collimated jet that is being
ejected from the very near 
black hole (along the tilted spin axis)
as the black hole orbits the center of mass would have
a spiral structure. A more detailed analysis would have to be done to see
if the wiggles in the radio jets with the $3\fd 0\pm0\fd 2$
periodicity
observed by 
Hjellming \&    Rupen 1995\markcite{hr95}
can be entirely explained simply by the orbital motion of the black hole.
In particular, one would have to consider the physical sizes of the observed
wiggles as well as the complicated physical processes taking place inside the
jets themselves.

If the tilt in the jet axis comes about because the spin axis of the black
hole is tilted, then one needs to explain the misalignment of the spin
axis of the black hole and that of the orbital plane.
The Monte Carlo simulations of supernovae kicks on neutron star
binaries by 
Brandt \& Podsiadlowski (1995)\markcite{bp95} 
predict that most of the massive stars in these neutron star
binaries would have inclined spin axes.  If the
work of 
Brandt \& Podsiadlowski (1995)\markcite{bp95}  
can be  extended to the
black hole systems, it may turn out that a $15.5\deg$ tilt
in the black hole's spin axis is not at all unusual.  
 
\subsection{Evolution of the System} 

Despite the pleasing agreement between single star evolutionary models
and the observed parameters of the secondary,
the evolutionary history of
GRO J1655-40 is clearly much more complex than the evolutionary
history of a single star (e.g.\  
Webbink 1992\markcite{we92}; 
van den Heuvel 1992\markcite{vdh92}; 
Webbink \& Kalogera 1994\markcite{wk94}).  
Webbink \& Kalogera (1994)\markcite{wk94}  
give several conditions for the successful formation of a low
mass X-ray binary, some of the conditions being:
(i)~the progenitor binary must have an extreme mass ratio
in order to drive the system to a common envelope
{\em and} the initial binary must be wide enough to survive the
common envelope stage; (ii)~the 
post-common-envelope binary must be big enough to accommodate the secondary
star within its Roche lobe {\em and} wide enough to allow the primary
to evolve to core collapse; and (iii)~the binary system must survive
the core collapse.  If the conditions stated by
Webbink \& Kalogera (1994)\markcite{wk94} 
are met and we are left with a
$7\,M_{\sun}$ black hole with a $2.3\,M_{\sun}$ main sequence companion
star in an 
orbit with a period of about 2.6 days, the subsequent binary evolution
will dominated by the evolution of the secondary star since the timescale
for the evolution of that $2.3\,M_{\sun}$ star is much shorter than
the timescale for the loss of orbital angular momentum 
(King, Kolb, \& Burderi 1996)\markcite{kkb96}.
The orbit is not likely to be left circular after the core
collapse, but tidal dissipation will
circularize the orbit before the secondary can fill its Roche lobe
(Webbink \& Kalogera 1994)\markcite{wk94}.  
Once the orbit
is circularized, the stellar evolution computations
we performed show that
on a timescale of order 690
million years, the companion star will evolve off the main sequence and
attain a radius, luminosity, and temperature comparable to
the secondary star of GRO J1655-40 that we see today.  
The radius the star attains is roughly the effective radius of the
Roche lobe of the secondary star in GRO J1655-40.

GRO J1655-40 stands out among the black hole binaries in that it has
a large space velocity (i.e.\ it has a large $\gamma$ velocity, see
Table \ref{sco1tab3} and 
Brandt, Podsiadlowski, \& Sigurdsson [1995\markcite{bps95}, 
hereafter BPS95]).  
BPS95\markcite{bps95} 
argued that GRO J1655-40 may
have acquired its velocity as the result of a kick caused
by an asymmetry during the initial collapse of the compact object
(as is thought to be the case with neutron stars---Brandt \
\& Podsiadlowski 1995\markcite{bp95}; 
Johnston 1996\markcite{jo96}). 
BPS95\markcite{bps95} 
proposed that the black hole in GRO J1655-40 
formed via an intermediate neutron stage, and
was then converted into a black hole by additional accretion from
the secondary star or through some kind of phase transition in the cooling
compact 
object.  This scenario predicts that the final black hole binary system
should have a large mass ratio and that the masses of the components
would be relatively small ($M_1\approx 3.6\,M_{\sun}$
and $M_2\approx 0.3\,M_{\sun}$).  However, our fits to the light curves show
that almost the opposite situation is true:  the mass ratio $Q$ is relatively
close to unity and the masses of the components are relatively
large ($M_1=7\,M_{\sun}$ and $M_2=2.3\,M_{\sun}$).   Therefore, if
the compact object passed through an intermediate neutron stage, it
would have had to accrete $\gtrsim 5\,M_{\sun}$ to attain its present
day mass, with the accreted matter presumably coming from the companion star.  
As we showed above, the position of the secondary star on the
Hertzsprung-Russell diagram is consistent with the track of a normal
$2.3\,M_{\sun}$ star halfway between the main sequence and the giant 
branch.  If the secondary star was once much more massive and it gave up most
of its mass to the compact primary, its evolutionary path on the
Hertzsprung-Russell diagram could be quite different than that of a normal
$2.3\,M_{\sun}$ star.
Thus, 
BPS95's\markcite{bps95} 
proposed scenario of the formation of the
black hole in GRO J1655-40 seems to be less viable
in view of the large amount of matter the
compact object must  accrete and the secondary star must give up
in order to attain their present-day masses.
If the compact object initially formed as a neutron star, it seems likely
that a significant amount of material from the supernova itself
must have fallen back, thus
converting the neutron star into a black hole shortly after its creation.
Note that the same supernova kick invoked by 
BPS95\markcite{bps95} 
to explain the
large velocity of GRO J1655-40 might also give rise to a substantial
tilt in the spin axis of the compact object as discussed in Section 6.2.

\subsection{Outburst Mechanism}

Recently, 
King et al.\ (1996)\markcite{kkb96}  
and 
van Paradijs (1996)\markcite{vp96} 
discussed
the transient behavior in some low mass X-ray binaries.  In general,
a system will be transient if the average mass transfer rate $\dot M$
is smaller than some critical value.  In the case of GRO J1655-40
where the main sequence lifetime of the secondary star is much shorter
than the timescale of the shrinkage of the orbit due to angular momentum
loss, the average mass loss rate is given by 
King et al.\ (1996)\markcite{kkb96} as
\begin{equation}
-\dot M_2 \approx 4\times10^{-10}P^{0.93}_dM^{1.47}_2\,M_{\sun}~{\rm yr}^{-1}
\label{mdot}
\end{equation}
where $P_d$ is the orbital period in units of days and where the units
of $M_2$ are solar masses.  For GRO J1655-40, $\dot M_2=3.4\times
10^{-9}\,M_{\sun}~{\rm yr}^{-1}$ $=2.16\times 10^{17}$~g~s$^{-1}$.
This average mass transfer rate is much larger
than the average mass transfer rates found in the other six transient
black hole systems, which all have rates around 
$10^{-10}\,M_{\sun}~{\rm yr}^{-1}$ 
(van Paradijs 1996\markcite{vp96}).  
For
X-ray heated accretion disks, the critical mass transfer rate is given
by
\begin{equation}
\dot M_{\rm crit}\approx 5\times 10^{-11}M^{2/3}_1P^{4/3}_3\,M_{\sun}~{\rm
yr}^{-1}
\label{crit}
\end{equation}
where the units of $M_1$ are solar masses and where
$P_3=P/(3\,{\rm hr})$ 
(King et al.\ 1996)\markcite{kkb96}.  
For GRO J1655-40, 
$\dot M_{\rm crit}=1.1\times 10^{-8}\,M_{\sun}~{\rm yr}^{-1}$.  Thus
GRO J1655-40 is {\em not} expected to be a persistent X-ray source since
$\dot M < \dot M_{\rm crit}$.  It is, however, interesting to note how close
GRO J1655-40 is to being a persistent X-ray source.  
van Paradijs (1996)\markcite{vp96} 
gives the following relation dividing systems
with stable and unstable mass transfer:
\begin{equation}
\log \bar{L_x}=35.8 + 1.07\log P({\rm hr})
\label{divid}
\end{equation}
where $\bar{L_x}$ is the time-averaged X-ray luminosity over one
outburst cycle.  If a system falls {\em below} the line defined
by Equation \ref{divid} in the $P-\bar{L_x}$~plane, 
then it will be a transient system.
If the energy generation rate is $0.2c^2$ per gram
of accreted matter
(van Paradijs 1996\markcite{vp96}), 
the averaged mass transfer rate of 
$=2.16\times 10^{17}$~g~s$^{-1}$ for GRO J1655-40
corresponds to an average X-ray
luminosity of $\bar{L_x}=3.88\times 10^{38}$~erg~s$^{-1}$.   This is
slightly smaller than the value of
$\bar{L_x}=5.31\times 10^{38}$~erg~s$^{-1}$ predicted from the relation
given by Equation (\ref{divid}).  The other six transient black hole systems
have average X-ray luminosities that are at least a factor of
ten less than the critical X-ray luminosity defined by Equation
\ref{divid} (see Figure 2 of 
van Paradijs 1996\markcite{vp96}), 
which suggests that
GRO J1655-40 is likely to have more frequent
X-ray outbursts than  the other six transient black hole systems.
Indeed, 
there were several additional hard X-ray outbursts after the initial hard
X-ray outburst observed July, 1994 
(Harmon et al.\ 1995a\markcite{ha95a}), 
and  the soft X-ray outburst observed in late April, 1996
(Remillard et al.\ 1996a)\markcite{re96} 
came less than one year after the sequence of
hard X-ray outbursts ended.

\section{Summary}

Using our database of GRO J1655-40 spectra, we have established
much improved values of the orbital period ($P=2\fd 62157\pm 0\fd 00015$),
the radial velocity semiamplitude  
($K_2=228.2\pm 2.2$~km~s$^{-1}$), the mass function 
($f(M)=3.24\pm 0.09\,M_{\sun}$), and MK spectral type (F3IV to F6IV).
Our photometry taken while the system was in true X-ray quiescence
shows that light from the distorted secondary star dominates, making
it possible to model the light curves in detail.  Our model of a distorted
secondary star plus a circular accretion disk provide excellent fits to the
light curves taken during true X-ray quiescence.  We can add the
effects of X-ray heating on the secondary star and fit the general
shape of the March 18-25, 1995 $V$ and $I$ light curves, taken during
a period of intense activity in hard X-rays.  The best values
of the orbital inclination angle $i$ and the mass ratio $Q$
($i=69.50\pm 0.08\deg$ and $Q=2.99\pm 0.08$) may be combined with the mass
function to give, for the first time, a reliable mass for a black
hole:  $M_1=7.02\pm 0.22\,M_{\sun}$.  The position of the secondary
star on the Hertzsprung-Russell diagram is well determined, and its
luminosity, temperature, and radius are all consistent with a $2.3\,M_{\sun}$
star $\approx 690$ million years into its evolution past the ZAMS.
The average mass accretion rate of the system is much larger than the
averaged accretion rates of the other transient black hole systems, putting
GRO J1655-40 much closer to the threshold where it would be a persistently
strong X-ray source.

\acknowledgements

We are grateful to Sydney Barnes and Dana Dinescu who assisted with the
observations, and to the staff of CTIO for the excellent support,
in particular Srs.\ Maurico Navarrete, Edgardo Cosgrove, Maurico
Fernandez, \& Luis
Gonzalez.  We thank the CTIO Time Allocation Committee for their flexible
scheduling of the spectroscopy run on the 1.5 meter telescope.
Alison Sills and Sukyong Yi kindly provided the stellar
evolution tracks for us.  Mr.\ Yi also gave us the digitized versions of
the standard filter response curves.  We thank
Nan Zhang for providing us with the BATSE daily flux
averages from late March, 1995.  We acknowledge useful discussions with
Craig Robinson, Jeffrey McClintock, Ronald Remillard, and Richard Wade.
Financial support for this work was provided by the National Science
Foundation through a National Young Investigator grant to C. Bailyn.

\appendix

\section{Appendix:  The Eclipsing Light Curve Code}

In this Appendix we describe in detail the code used to model the eclipsing
light curve of GRO J1655-40.  This program is a modified version
of code first written by Yoram Avni 
(Avni \& Bahcall 1975\markcite{ab75}; 
Avni 1978\markcite{av78};
see also 
McClintock \& Remillard 1990\markcite{mr90}).  
We will outline here in detail
the basic input physics and approximations used, not to take credit
for the work of Avni and others, but to tell the readers exactly what
the code does and how it does it.  

\subsection{The Potential}
 
Consider a binary system consisting of a visible star
of mass $M_2$  and a compact object of mass $M_1$,
where the visible star orbits in a circular orbit
with a Keplerian angular velocity $\omega_k$.  Assume the visible star is
also rotating with an angular velocity $\omega_2$.  Following the
notation in 
Avni (1978)\markcite{av78}, 
we define a rectangular coordinate system
with the origin at the center of the visible star, and
with the $X$-axis pointing towards the compact object, and the
$Z$-axis in the direction of $\vec \omega_k$.  This coordinate system
rotates with the visible star.  
The potential can be written as
(Avni 1978\markcite{av78})
\begin{eqnarray}
\Psi(X,Y,Z)  &  = &  - {GM_2\over D}\left[{1\over r_2}+{Q\over r_1}-Qx 
                                    \right. \nonumber           \\ 
             &  & +  \left. {\Omega^2(x^2+y^2)(1+Q)\over 2}\right]         
\label{neweq1}
\end{eqnarray}
where $D$ is the separation of the two stars, $Q=M_1/M_2$,
$\Omega=\omega_2/\omega_k$, $r_1$ and $r_2$
are the distance to the centers of the two stars in units of D, 
and where $x$ and $y$, are
$X$ and $Y$ in units of $D$.
If $\Omega=1$ (the visible star is in synchronous rotation), the
potential given by Equation (\ref{neweq1}) reduces to the standard
Roche potential, and the star can be in hydrostatic equilibrium in 
the rotating frame 
(Avni 1978\markcite{av78}).  
The degree to which the star fills
its Roche lobe must be specified---it is usually taken to be 100\%.

In practice, it is convenient to adopt units of mass and distance such
that the $GM_1/D=1$.  Typically, the star is assumed to be in
synchronous rotation, so that $\Omega=1$, and it is assumed
that it completely fills its Roche lobe.  Thus the value of the
mass ratio
$Q$ uniquely determines the function for the potential, and
hence the geometry of the Roche surface.  Once the value of $Q$ is
specified, the visible star is divided into
 $N_\phi$ grid points in longitude $\phi$, where the
points are spaced equally in $\cos\phi$,  and
$4N_\theta$ grid points in the co-latitude $\theta$, where the points
are spaced equally in the angle $\theta$.  The value of the potential
$\Psi$ and its derivatives are then computed for each point.

\subsection{Photometric Parameters}\label{secvZ}
 
von Zeipel's Theorem (1924)\markcite{vz24} 
provides a relationship between the
local gravity and the local emergent flux in a tidally distorted star.
Proofs of 
von Zeipel's Theorem may be found in
Kopal \& Kitamura (1968)\markcite{kk68} 
and 
Avni (1978\markcite{av78}).  
Using von Zeipel's Theorem, 
one can show
that the light emitted from every point on the photosphere of the star
is the same as the light emitted from a plane-parallel
atmosphere characterized by the local values of the temperature
$T_e$ and gravity $g$:
\begin{equation}
T_e^4\propto g.
\label{neweq6}
\end{equation}
As a consequence of von Zeipel's Theorem, the temperature at any point on
the star is given by
\begin{equation}
{T(x,y,z)\over T_{\rm pole}}=\left[{g(x,y,z)\over g_{\rm pole}}
\right]^\beta,
\label{neweq8}
\end{equation}
where $T_{\rm pole}$ and $g_{\rm pole}$ are the temperature and gravity
at the pole of the star (i.e.\ the point on the
surface of the star where the positive $Z$-axis emerges).
The ``gravity darkening exponent'' $\beta$ has two values: 0.25 for stars
with radiative atmospheres
as shown by 
von Zeipel (1924)\markcite{vz24}, 
and 0.08 for stars with fully convective envelopes 
(Lucy 1967)\markcite{lu67}.  
Thus, to specify the temperature $T(x,y,z)$
at every point on the star, one must input the value of $T_{\rm pole}$.
This input temperature is usually taken to be the effective temperature
of a field star with a similar spectral type as the star to be modelled.
In our model,
the surface gravity on each point on the star can be computed from the
derivatives of the potential:
\begin{equation}
g(x,y,z)=-\left[\left(\partial\Psi\over\partial x\right)^2 +
\left(\partial\Psi\over\partial y\right)^2 +
\left(\partial\Psi\over\partial z\right)^2\right]^{1/2}
\label{neweq7}
\end{equation}
e.g.\ (Zhang, Robinson, \& Stover 1986\markcite{zrs86}).  

Once the temperature is computed for each surface point, 
Planck's function
is used to approximate the relationship between the temperature and the
monochromatic intensity
\begin{equation}
I(\lambda)\propto\left[\exp\left({hc\over k\lambda T_e}\right)-1\right]^{-1},
\label{neweq10}
\end{equation}
where $h$, $c$, and $k$ are the usual physical constants.
The star appears darker near its limb, a phenomenon referred to as
limb darkening.
The code presently uses a standard linear limb darkening law expressed as
\begin{equation}
I(\lambda)\propto 1-u(\lambda)+u(\lambda)\cos\mu,
\label{neweq11}
\end{equation}
where $\mu$ is the angle between the surface normal and the line of
sight.  The values of the coefficient $u(\lambda)$ are taken from standard
tables computed from model atmospheres 
(e.g.\ Al-Naimiy 1978)\markcite{al78}.
The reader is referred to 
Kopal (1959)\markcite{ko59}, 
Kopal \& Kitamura (1968)\markcite{kk68}, 
and
Avni (1978\markcite{av78}) 
for more discussions on these approximations.
 
\subsection{Integration of the Flux from the Star}
 
Let $L(\lambda)$ 
be the radiation emitted by the star
at the wavelength $\lambda$  
as seen at a great distance.  If $I(\lambda,x,y,z)$ is the intensity
of the light at a surface point, the total observed flux is
given by
\begin{equation}
L(\lambda)=\int I(\lambda,x,y,z)\cos\gamma\, ds,
\label{neweq12}
\end{equation}
where $\gamma$ is the angle of foreshortening, $ds$ is the surface
element, and where the integration is to be done over the
entire visible surface of the star
(Kopal \& Kitamura 1968\markcite{kk68}).
 
To carry out the numerical integration of the flux, some quantities
need to be defined.  First, each point on the surface has direction cosines:
\begin{eqnarray}
    \ell_x &=& \cos\phi            \nonumber   \\
    \ell_y &=& \sin\phi\cos\theta   \nonumber    \\
    \ell_z &=& \sin\phi\sin\theta.   
\label{neweq14}
\end{eqnarray}
The element of the surface area $dS(x,y,z)$ is given by
\begin{equation}
dS(x,y,z)={R^2\Delta\phi\,\Delta\theta\over\sigma(x,y,z)}
\label{neweq15}
\end{equation}
where $R^2=x^2+y^2+z^2$,\, $\Delta\phi$\, and $\Delta\theta$\, are the
grid sizes in longitude and co-latitude, respectively, and where
\begin{eqnarray}
\sigma(x,y,z) &=& -{\ell_x\over g(x,y,z)}\left(
          {\partial\Psi(x,y,z)\over \partial x}\right) \nonumber \\ 
 & &  -{\ell_y\over g(x,y,z)}\left(
          {\partial\Psi(x,y,z)\over \partial y}\right) \nonumber \\
 & &  -{\ell_z\over g(x,y,z)}\left(
          {\partial\Psi(x,y,z)\over \partial z}\right).
\label{neweq16}
\end{eqnarray}
Next, let $\Theta$ be the orbital phase of the observation (where $\Theta=0$
corresponds to the time of the closest approach of the visible star) and
let $i$ be the inclination of the orbit ($i=90\deg$ for an orbit seen edge-on).
The foreshortening $\Gamma(x,y,z)$
of a particular point on the star depends on the phase
of the observation, the inclination, and on the $(x,y,z)$ coordinates
of the point:
\begin{eqnarray}
\Gamma(x,y,z) &=& -{1\over g(x,y,z)}\left(
  {\sin i\cos\Theta}{\partial\Psi(x,y,z)\over \partial x}\right)\nonumber \\
& &  +{1\over g(x,y,z)}\left(
  {\sin i\sin\Theta}{\partial\Psi(x,y,z)\over \partial y}\right)\nonumber \\
& &  +{1\over g(x,y,z)}\left(
  {\cos i}{\partial\Psi(x,y,z)\over \partial z}\right).
\label{neweq17}
\end{eqnarray}
If the ``projection factor'' $\Gamma(x,y,z)<0$, then that particular
point is not visible.  
At each phase, the flux elements from the visible points 
(i.e.\ those with $\Gamma(x,y,z)>0$) are simply summed up:
\begin{equation}
L_{\rm star}(\Theta,\lambda) 
= \sum_{i=1}^{N_{\phi}}\sum_{j=1}^{4N_{\theta}}I(\lambda,x,y,z)
\Gamma(x,y,z)dS(x,y,z),
\label{neweq18}
\end{equation}
where each pair of $(i,j)$ indices are associated with a specific
$(x,y,z)$ point on the star.
In practice, the above sum in Equation (\ref{neweq18}) 
converges effectively for $N_{\phi}\geq 40$
and $N_{\theta}\geq 14$.
 
So far, Equations (\ref{neweq1}) through (\ref{neweq18}) describe
the basic input physics and mathematics of the original Avni
code 
(Avni \& Bahcall 1975\markcite{ab75}; 
Avni 1978\markcite{av78}).  
This code models the light
curve due to a single Roche lobe filling star.  
Extra sources of light 
such as light due to X-ray heating effects are not taken into account
in the Avni code.  
To summarize the model so far, the user specifies
the degree to which the star fills its Roche lobe (usually 100\%), the value
of the mass ratio $Q$, and the rate of the star's rotation
(usually $\Omega=1$ for synchronous rotation).  These three quantities
define the shape of the surface of the star.  Then the user must specify
the polar temperature of the star $T_{\rm pole}$, the gravity darkening
exponent
$\beta$ (either 0.25 or 0.08 depending on whether the star
has a radiative envelope or a convective envelope),
and the linearized
limb darkening coefficient $u(\lambda)$ appropriate for the star in
question in order to compute the temperatures over the surface of the
star.   Once the temperatures are known, the Planck function and
a linearized limb darkening law are used to compute the intensities
over the surface.  Finally, after the orbital phase and the inclination
of the orbit are specified, the total observed flux can be computed.
 
\subsection{The Addition of an Accretion Disk}

To make the basic Avni code more realistic, we added an accretion disk 
to the code. Following 
Zhang et al.\ (1986)\markcite{zrs86}, 
the disk is flattened cylinder centered on the
(invisible) compact object.  The plane of the orbit bisects the disk
in the $z$-direction. 
The disk is assumed to be
completely optically thick, so that we can only see light
from its surface and so that anything behind the disk is completely
eclipsed. The face of the
disk is divided up into several grids equally spaced in the polar
coordinates $(r,\alpha)$.  The outer radius of the disk $r_d$
is scaled to a user specified fraction 
of the effective Roche lobe radius of the primary
and the inner radius $r_i$ is set to a very small number (typically 0.005).
Eggleton's (1983)\markcite{egg83}
approximation is used to compute the effective Roche lobe radius.
The thickness of the disk at the outer edge is $z_d=2r_d\tan\beta_{\rm rim}$. 
With the exception of user-defined ``hot-spots'', the temperature of the
disk does not vary with the azimuth angle.  The radial distribution
of the temperature across the face of the disk is given by
\begin{equation}
T(r)=T_{\rm disk}(r/r_d)^\xi,
\label{neweq21}
\end{equation}
where $T_{\rm disk}$\, is the user specified temperature at the outer
edge of the disk.
For a steady-state, optically thick, viscous accretion disk, the
power law exponent $\xi=-0.75$ 
(Pringle 1981)\markcite{pr81}.
As before, Planck's function is used to approximate the relationship
between the temperature of a point on the surface of the disk and the
monochromatic intensity $I_{\rm disk}(\lambda,r,\alpha)$ of that point.

Recently 
Diaz, Wade, \& Hubeny (1996)\markcite{dwh96} 
discussed the importance of
including corrections for limb darkening in models of disk spectra.
Limb darkening is important for disks since the limb darkening
corrections depend on the temperature (and hence the radius in our model)
and on the effective wavelength.  Therefore the slope of a blackbody
disk's spectrum will be in error if no limb darkening corrections
are made.  An important result of the work of 
Diaz et al.\ (1996)\markcite{dwh96}  
is that the limb darkening law in the optically thick rings of their accretion
disk models is very similar to the limb darkening law in a stellar atmosphere.
As a result, one may use the same linearized limb darkening coefficients
for disks that one uses for stars.  We therefore have
\begin{eqnarray}
I_{\rm disk}(\lambda,r,\alpha) & \propto & 1-u_{\rm disk}(\lambda,T(r))
                          \nonumber  \\
            &    & + u_{\rm disk}(\lambda,T(r))\cos i
\label{darkdisk}
\end{eqnarray}
where the values of $u_{\rm disk}(\lambda,T(r))$ are interpolated from
tables given by 
Wade \& Rucinski (1985)\markcite{wr85}.

The foreshortening angle of the normal of a surface element
on the face of the disk is simply $\cos i$.  In the absence of
eclipses, the observed flux from the face of the accretion disk is:
\begin{equation}
L_{\rm disk}(\Theta,\lambda)=\sum_{i=1}^{N_r}\sum_{j=1}^{N_\alpha}
I_{\rm disk}(\lambda,r,\alpha)(\cos i)\,r\Delta r\,\Delta\alpha,
\label{neweq22}
\end{equation}
where $N_r$ and $N_{\alpha}$ are the number of grid points in radius
and azimuth, respectively, where $\Delta r$\, and
$\Delta\alpha$\, are the grid spacings in
radius and azimuth, and where each pair of $(i,j)$ indices are associated
with a specific $(r,\alpha)$ point on the disk.
 
To handle the situations where the disk has a substantial thickness, light
from the rim is also accounted for.  The rim is divided into 11 grid points
in the $z$ direction, and $N_\alpha$ grid points in azimuth.  The temperature
of the rim is $T_{\rm disk}$, and the intensities of each rim point are
found using Planck's function. 
If the azimuth angle $\alpha$ is measured
from the $X$-axis, then the foreshortening factor 
$\Gamma_{\rm rim}(\alpha)$\, of a point on the rim at the
orbital phase $\Theta$ is
\begin{equation}
\Gamma_{\rm rim}(\alpha)=\sin i\,\cos(\alpha+\Theta).
\label{neweq23}
\end{equation}
If $\Gamma_{\rm rim}(\alpha)<0$, the point is hidden.  
A limb darkening correction is made for the edge surface elements based on
the value of $\Gamma_{\rm rim}(\alpha)$:
\begin{eqnarray}
I_{\rm rim}(\lambda,\alpha) & \propto & 1-u_{\rm disk}(\lambda,T_{\rm disk})
                                      \nonumber \\
    &    & + u_{\rm disk}(\lambda,T_{\rm disk})\Gamma_{\rm rim}(\alpha)
\label{darkdisk1}
\end{eqnarray}
The total flux
from the rim is found by summing the intensities of all visible 
points:
\begin{equation}
L_{\rm rim}(\Theta,\lambda)=\sum_{i=1}^{11}\sum_{j=1}^{N_\alpha}
I_{\rm rim}(\lambda,\alpha)\Gamma_{\rm rim}(\alpha)
\Delta z\,\Delta\alpha,
\label{neweq24}
\end{equation}
where $\Delta z$\, is the step size in the $Z$ direction,
and where each pair of $(i,j)$ indices are associated with a specific
$(z,\alpha)$ point on the rim.
 
Hot spots can be added to the disk.  The spots are confined to radii
$r_{\rm cut} \leq r \leq r_d$ and azimuth
angles $\alpha_1^i\leq \alpha
\leq \alpha_2^i$, where at the moment $i=0$ (no spots), 1 or 2.
The temperature of the spot $T_{\rm spot}$ is added to the
temperature $T(r,\alpha)$ of each point on the
disk rim and face that lies inside the spot boundaries.  The intensities of
each point
are computed as before and the integrations are carried out as
before (e.g.\ Equations (\ref{neweq22}) and (\ref{neweq24})).
 
In the absence of eclipses, the total observed flux from the star and
the disk is 
\begin{equation}
L_{\rm total}(\Theta,\lambda)=L_{\rm star}(\Theta,\lambda)+
L_{\rm disk}(\Theta,\lambda)+L_{\rm rim}(\Theta,\lambda).
\label{neweq25}
\end{equation}
 
\subsection{The Addition of Eclipses}

In the case where the inclination angle $i$ is large enough to allow for
eclipses, the procedure for computing the light curves must be modified
slightly.  First, the intensities of each point on the star, disk face, and
disk rim are computed from the list of input parameters.  We can determine
which body is in front of the other body from the
orbital phase $\Theta$---the star will obviously be closer than the disk
for $0\deg\leq\Theta\leq 90\deg$ or for $270\deg
\leq\Theta\leq 360\deg$ and the other way around for
$90\deg\leq\Theta\leq 270\deg$.  The integration of the flux elements
over the visible surface of the object in front is carried out as before.
Then,  all points on the edge of the body in front
(i.e.\ its horizon) are determined and are projected
onto the plane of the sky.  The horizon of the eclipsing body defines
some polygon on the plane of the sky.  Then, as the integration of the
flux elements of the body in back is carried out, each potentially
visible point is projected
onto the plane of the sky and
checked to see if it falls inside the horizon of the eclipsing body.  If
the point is eclipsed, it is not included in the integration.
Once all of the visible points are included in the integrations, the
total observed flux is found as before
(Equation (\ref{neweq25})).
 
\subsection{The Addition of X-ray Heating}
 
If the secondary star intercepts a large amount of flux from the other
body (either X-ray heating from the vicinity of the compact object or
optical light from the disk), there may be a large amount of reprocessed
radiation.  The irradiating flux can strongly alter the distribution
of temperatures across the face of the star.  A rigorous treatment
of these ``reflection effects'' requires difficult and time-consuming
computations (e.g.\ 
Kopal 1959\markcite{ko59}; 
Wilson \& Devinney 1971\markcite{wd71}).  
Following
Zhang et al.\ (1986)\markcite{zrs86}, 
we will use a simplified treatment of the heating
effects.  We will consider only heating of the star by X-ray heating, 
and not the heating of the disk by the star.  For simplicity,
all of the irradiating flux $F_{\rm irr}$ due to the
X-ray heating is assumed
to come from the center of the compact object. 
 
The {\em total} energy in X-rays intercepted by the Roche
lobe filling secondary star does not
depend on the scale of the orbit since the surface area 
of Roche lobe filling star
facing the X-ray source is proportional to the square of the orbital
separation.  However,
the change in the star's atmosphere caused by the X-ray heating
{\em does} depends
on the size of the orbit.  As the distance between the star and the
X-ray source is increased, each square centimeter on the star's surface
receives less energy in X-rays.  The change in the local temperature
depends on the amount of absorbed X-ray flux per unit area.
So to find the magnitude of the irradiating flux seen by 
each point on the secondary,
the X-ray luminosity of the X-ray source $L_x$ (in ergs~s$^{-1}$)
and the physical size of the orbit  must be specified.
The code uses
the mass function $f(M)$ and the orbital
period $P$ to find the scale of the orbit.  
 
The distance $d$ (in units of the orbital separation) between each point
on the face of the star and the X-ray source can be found
\begin{equation}
d(x,y,z)=(1+z^2+R^2-2R\ell_x),
\label{neweq26}
\end{equation}
where $\ell_x$ is one of the direction cosines (Equation (\ref{neweq14})),
and where $R^2=x^2+y^2+z^2$.  The foreshortening factor
$\Gamma_{\rm X-ray}(x,y,z)$ is given by
\begin{equation}
\Gamma_{\rm X-ray}(x,y,z)={x(1-x)-y^2+z^2\over Rd},
\label{neweq27}
\end{equation}
where $\Gamma_{\rm X-ray}(x,y,z)<0$ indicates that the point on the
star cannot see the X-ray source.
The amount of X-ray flux each surface element intercepts is
\begin{equation}
F_{\rm irr}(x,y,z)=\left({L_x\over 4\pi a^2d(x,y,z)^2}\right)
\Gamma_{\rm X-ray}(x,y,z).
\label{newew28}
\end{equation}
The modified temperature of each surface point that can see the X-ray source
is given by
\begin{equation}
T_{\rm X-ray}^4(x,y,z)=T_{\rm pole}^4\left({g(x,y,z)
\over g_{\rm pole}(x,y,z)}\right)^{4\beta}+{WF_{\rm irr}\over \sigma}
\label{neweq29}
\end{equation}
where $W$ is the albedo, and $\sigma$ is the Stefan-Boltzmann constant
(Zhang et al.\ 1986)\markcite{zrs86}.  
The albedo $W$ is not very well determined, and
it is usually set to 0.5 (see 
Zhang et al.\ 1986\markcite{zrs86}).  
 
In the cases there the disk has a substantial thickness, some points
on the face of the star may be below the rim of the disk as seen
from the compact object.  These points would not receive any
X-ray flux from the compact object and hence their temperatures would not
change.  It is easy to find the angle above the plane that a point on
the face of the star has (as seen from the compact object):
\begin{equation}
\Phi(x,y,z)=\arctan\left({z\over (1-x)\cos\eta}\right)  
\label{neweq30}
\end{equation}
where $\eta=\arctan(y/x)$.
This angle $\Phi(x,y,z)$ 
is compared to the angle of the rim $\beta_{\rm rim}$
to see whether or not the point on
the star is blocked by the disk.  If $\beta_{\rm rim}>\Phi(x,y,z)$,
the point cannot see the X-ray source.
 
Once each point on the face of the star has been assigned a new temperature
based on the amount of X-ray flux it sees, the intensities 
and the fluxes are found
following the procedures outlined above.  The net result is that
the points on the star facing the X-ray source are brighter than they
would otherwise be because the temperatures were raised.

\subsection{Filter Response Curves}

We compute the total flux $L_{\rm total}(\Theta,\lambda)$
at each phase $\Theta$ 
for several values of the effective wavelength
$\lambda$ using limb darkening coefficients $u(\lambda)$ interpolated
from tables given by 
Wade \& Rucinski (1985)\markcite{wr85}.  
This model ``spectrum'' is
integrated with the standard $UBVRI$ filter response curves
(Bessell 1990\markcite{bess90}) 
to produce
model $UBVRI$ magnitudes.  In this way we can perform simultaneous
fits to light curves in several colors.  In the case of GRO J1655-40,
there is a large amount of reddening in the light curves.  Rather than
attempting to include the reddening correction in the fits, we instead
fit the models to the normalized $BVRI$ data.  In this way, the error of
$E(B-V)$ does not enter into the analysis.  We do, however, lose
the absolute color information contained in the model.

\subsection{Light Curve Fitting and Error Analysis}

We first compute a grid of models and $\chi^2$ fits to the data using a wide
range of parameters spaced in uniform steps.  Then several sets of
parameters that produce low $\chi^2$ values are chosen and
are used as starting points for an optimization routine based
on the GRIDLS program given in 
Bevington (1969)\markcite{be69}.  
This ``grid search''
method was picked because it is easy to implement and it does not require
the computation of derivatives.  After several runs with different
starting points, the set of parameters that gives the lowest
overall $\chi^2$ fit to the data is chosen.  By using several different
starting points for the optimization routine, we can be reasonably sure
that we have found the global minimum of $\chi^2$, rather than a
local minimum.

We use three methods to estimate the statistical errors on the fitted
parameters.  The first one is a ``bootstrap'' method 
(Press et al.\ 1992)\markcite{ptvf92}.  
Artificial
data sets are constructed from the original data set and the exact
fitting procedures which were used on the original data
are used on the artificial data.  The $1\sigma$ errors on the parameters
are then derived from the distributions of the fitted parameters from
the many artificial data sets.  The second technique we use is outlined 
by 
Zhang et al.\ (1986)\markcite{zrs86}.
Basically the $1\sigma$
error on a fitted parameter $a_0$ is found by fixing the value of that
parameter at a new value $a^{\prime}_0=a_0+\delta$ and finding the values
of the rest of the parameters $a_1$ through
$a_n$ that minimize $\chi^2$.  The
probable error of $a_0$ is related to the change in  $\chi^2$ and to $\delta$.
We typically use two or three different values of $\delta$ 
(1\%, 2\%, and 3\% of the value of the parameter) and average the resulting
values of $\sigma$.
Finally, the GRIDLS program given in 
Bevington (1969)\markcite{be69} 
also 
provides error estimates based on the shape of the chi-square hypersurface
near the minimum.  
We used primarily the errors derived from the bootstrap method.
We stress here that the errors estimated are
only internal statistical errors, and not systematic errors which we have
no easy way to compute.

\end{document}